\documentclass[twocolumn]{aastex631}

\usepackage{amsmath}
\usepackage{bm}
\usepackage{gensymb}
\usepackage{graphicx}
\usepackage{hyperref}
\usepackage{multirow}
\usepackage{soul}
\usepackage[nameinlink]{cleveref}
\usepackage[multiple]{footmisc}
\crefname{paragraph}{\S}{\S\S} % default is {paragraph}{paragraphs}
\shorttitle{MOA-2007-BLG-192} 
\shortauthors{Terry et al.}

\begin{document}

%\title{Hubble and Keck Follow-Up Observations Show a Low-Mass M Dwarf is Hosting the Microlensing Planet MOA-2007-BLG-192Lb}

%Better title?:
\title{Unveiling MOA-2007-BLG-192: An M Dwarf Hosting a Likely Super-Earth}

\author[0000-0002-5029-3257]{Sean K. Terry}
\affiliation{Department of Astronomy, University of Maryland, College Park, MD 20742, USA}
\affiliation{Code 667, NASA Goddard Space Flight Center, Greenbelt, MD 20771, USA}

\author[0000-0003-0014-3354]{Jean-Philippe Beaulieu}
\affiliation{Sorbonne Universit\'e, CNRS, Institut d’Astrophysique de Paris, IAP, F-75014 Paris, France}
\affiliation{School of Natural Sciences, University of Tasmania, Private Bag 37 Hobart, Tasmania, 7001, Australia}

\author[0000-0001-8043-8413]{David P. Bennett}
\affiliation{Department of Astronomy, University of Maryland, College Park, MD 20742, USA}
\affiliation{Code 667, NASA Goddard Space Flight Center, Greenbelt, MD 20771, USA}

\author{Euan Hamdorf}
\affiliation{School of Natural Sciences, University of Tasmania, Private Bag 37 Hobart, Tasmania, 7001, Australia}

\author{Aparna Bhattacharya}
\affiliation{Department of Astronomy, University of Maryland, College Park, MD 20742, USA}
\affiliation{Code 667, NASA Goddard Space Flight Center, Greenbelt, MD 20771, USA}

\author{Viveka Chaudhry}
\affiliation{Sidwell Friends School, Washington, DC 20016, USA}

\author[0000-0003-0303-3855]{Andrew A. Cole}
\affiliation{School of Natural Sciences, University of Tasmania, Private Bag 37 Hobart, Tasmania, 7001, Australia}

\author[0000-0003-2302-9562]{Naoki Koshimoto}
\affiliation{Department of Earth and Space Science, Graduate School of Science, Osaka University, Osaka, 560-0043, Japan}

\author[0000-0003-2861-3995]{Jay Anderson}
\affiliation{Space Telescope Science Institute, 3700 San Martin Drive, Baltimore, MD 21218, USA}

\author{Etienne Bachelet}
\affiliation{IPAC, Caltech, Pasadena, CA 91125, USA}

\author[0000-0001-5860-1157]{Joshua W. Blackman}
\affiliation{Physikalisches Institut, Universit{\"a}t Bern, Gesellschaftsstrasse 6, CH-3012 Bern, Switzerland}

\author[0000-0002-8131-8891]{Ian A. Bond}
\affiliation{School of Mathematical and Computational Sciences, Massey University, Auckland 0632, New Zealand}

\author[0000-0001-9611-0009]{Jessica R. Lu}
\affiliation{Department of Astronomy, University of California Berkeley, Berkeley, CA 94720, USA}

\author[0000-0002-7901-7213]{Jean Baptiste Marquette}
\affiliation{Laboratoire d'Astrophysique de Bordeaux, CNRS, B18N, all{\'e}e Geoffroy Saint-Hilaire, Pessac, France}
\affiliation{Sorbonne Universit\'e, CNRS, Institut d’Astrophysique de Paris, IAP, F-75014 Paris, France}

\author[0000-0003-2388-4534]{Cl\'ement Ranc}
\affiliation{Sorbonne Universit\'e, CNRS, Institut d’Astrophysique de Paris, IAP, F-75014 Paris, France}

\author[0000-0002-1530-4870]{Natalia E. Rektsini}
\affiliation{School of Natural Sciences, University of Tasmania, Private Bag 37 Hobart, Tasmania, 70001, Australia}
\affiliation{Sorbonne Universit\'e, CNRS, Institut d’Astrophysique de Paris, IAP, F-75014 Paris, France}

\author[0000-0001-6008-1955]{Kailash Sahu}
\affiliation{Space Telescope Science Institute, 3700 San Martin Drive, Baltimore, MD 21218, USA}

\author[0000-0002-9881-4760]{Aikaterini Vandorou}
\affiliation{Department of Astronomy, University of Maryland, College Park, MD 20742, USA}
\affiliation{Code 667, NASA Goddard Space Flight Center, Greenbelt, MD 20771, USA}

\correspondingauthor{S. K. Terry}
\email{skterry@umd.edu}



\begin{abstract}

\noindent We present an analysis of high angular resolution images of the microlensing target MOA-2007-BLG-192 using Keck adaptive optics and the \textit{Hubble Space Telescope}. The planetary host star is robustly detected as it separates from the background source star in nearly all of the Keck and \textit{Hubble} data. The amplitude and direction of the lens-source separation allows us to break a degeneracy related to the microlensing parallax and source radius crossing time. Thus, we are able to reduce the number of possible binary lens solutions by a factor of ${\sim}2$, demonstrating the power of high angular resolution follow-up imaging for events with sparse light curve coverage. Following \citep{bennett:2023a}, we apply constraints from the high resolution imaging on the light curve modeling to find host star and planet masses of $M_{\rm host} = 0.28 \pm 0.04M_{\sun}$ and $m_p = 12.49^{+65.47}_{-8.03}M_{\Earth}$ at a distance from Earth of $D_L = 2.16 \pm 0.30\,$kpc. This work illustrates the necessity for the \textit{Nancy Grace Roman Galactic Exoplanet Survey} (\textit{RGES}) to use its own high resolution imaging to inform light curve modeling for microlensing planets that the mission discovers.
\end{abstract}



\section{Introduction} \label{sec:intro}

\indent Since the early 1990's, surveys of the galactic bulge have searched for variations in the brightness of background stars (sources) induced by the gravitational field of foreground objects (lenses). The number of lensing events detected has dramatically increased from a few dozen per year in the 1990's \citep{udalski:1994a, alcock:1996a} to thousands per year currently. At present there are three primary ground-based surveys that contribute to these lensing event detections: OGLE \citep{udalski:2015a}, MOA \citep{bond:2001a}, and KMTNet \citep{kim:2016a}. NASA's Nancy Grace Roman Space Telescope (\textit{Roman}) is scheduled to launch in the next several years and will conduct the \textit{Roman} Galactic Bulge Time Domain Survey (\textit{GBTDS}) {\citep{Gaudi:2022a}}. As part of this bulge survey, the Roman Galactic Exoplanet Survey (\textit{RGES}) will be the first dedicated space-based gravitational microlensing survey and is expected to detect over 30,000 microlensing events and over 1400 bound exoplanets during its five-year survey \citep{penny19}. This mission will complement previous large statistical studies of transiting planets from missions like \textit{Kepler}/\textit{TESS} and radial velocity (RV) planets from many ground-based RV surveys. The \textit{GBTDS} is also expected to discover free-floating planets that do not orbit any host star (\cite{johnson:2020a, sumi:2023a}, Johnson et al. in prep).\\
\indent As of this writing, microlensing has detected $\sim$200 planets at distances up to the Galactic Bulge \footnote{https://exoplanetarchive.ipac.caltech.edu/}. As for most transient phenomena, one limitation of this method for fully characterizing microlensing systems is the cadence at which the photometric data is obtained by the dedicated ground-based surveys. An effective way to increase the sampling for a given microlensing event is to issue a public alert so that observatories around the world can observe ongoing events as a `follow-up' network of telescopes. MOA-2007-BLG-192 was the first planetary microlensing event detected without follow-up observations from other observatories. The initial analysis reported a low-mass planet orbiting a very-low mass host star or brown dwarf \citep{bennett:2008b}. Due to the lack of follow-up network data for this microlensing event, there are significant gaps in the photometric light curve coverage, which leads to uncertainties in the derived lens system parameters. There are also additional degeneracies in the interpretation of this lens system that arise from the possible planet-star separations and microlensing parallax. The details of these degeneracies are discussed further in Section \ref{subsec:bennett-work}.\\
\indent One way to mitigate some of these degeneracies is by resolving the source and lens independently with high angular resolution imaging (i.e. \textit{Hubble Space Telescope} (HST), Keck telescopes, Subaru telescope) several years after peak magnification \citep{bennett:2006a,bennett:2007a}. This high angular resolution imaging can enable measurements of the lens-source separation, relative proper motion, and lens flux which can then be used with mass-luminosity relations \citep{henry:1993a,henry:1999a,delfosse:2000a} to calculate a direct mass for the host.\\
\indent This current analysis is part of the NASA Keck Key Strategic Mission Support (KSMS) program,
``Development of the WFIRST Exoplanet Mass Measurement Method" \citep{bennett_KSMS}, which is a pathfinder project for the \textit{Nancy Grace Roman Space Telescope}  (formerly known as \textit{WFIRST}) \citep{spergel:2015a}. For several years now, the KSMS program has measured the masses of many microlensing host stars and their companions \citep{bhattacharya:2018a, vandorou:2019a, bennett:2020a, blackman:2021a, terry:2021a, terry:2022a}, all of which are included in one of the most complete statistical studies of the microlensing exoplanet mass ratio function \citep{suzuki:2016a, suzuki:2018a}. This statistical study shows a break and likely peak in the mass-ratio function for wide-orbit planets at about a Neptune mass which is at odds with the runaway gas accretion scenario of the leading core accretion theory of planet formation \citep{lissauer93,pollack96}, which predicts a planet desert at sub-Saturn masses \citep{ida:2004a} for gas giants at wide orbits. \\
\indent This paper is organized as follows: In Section \ref{sec:event} we present the light curve re-analysis of MOA-2007-BLG-192 and explain the challenges in the modeling posed by lack of photometric coverage and degeneracies. In Section \ref{sec:follow-up}, we describe the high angular resolution \textit{HST} and Keck adaptive optics (AO) observations and analysis. Section \ref{sec:prop-motion} details our direct measurement of the lens system flux and lens-source separation in the Keck and \textit{HST} data which allows us to reduce the total number of binary lens solutions. Section \ref{sec:lens-properties} describes the newly derived lens system properties from the light curve modeling that incorporates the high resolution imaging results. Finally, we discuss the overall results and conclude the paper in Section \ref{sec:conclusion}.

%----------------------------------------------------------Original Event------------------------------------------------------------------------------------------------------------

\section{Prior Studies of the Microlensing Event MOA-2007-BLG-192} \label{sec:event}

\subsection{Fitting the microlensing light curve} \label{subsec:bennett-work}

MOA-2007-BLG-192 (hereafter MB07192), located at RA (J2000) $=$ 18:08:03.80, DEC (J2000) $=$ $-$27:09:00.27 and Galactic coordinates $(l,b)=(4.03\degree, -3.39\degree)$ was first alerted by MOA on May 24, 2007. Due to the faintness of the source and poor weather at the MOA telescope, the event was not alerted until the day that the planetary deviation was observed in the light curve. 

\begin{figure*}
%\figurenum{1}
\includegraphics[width=0.8\linewidth]{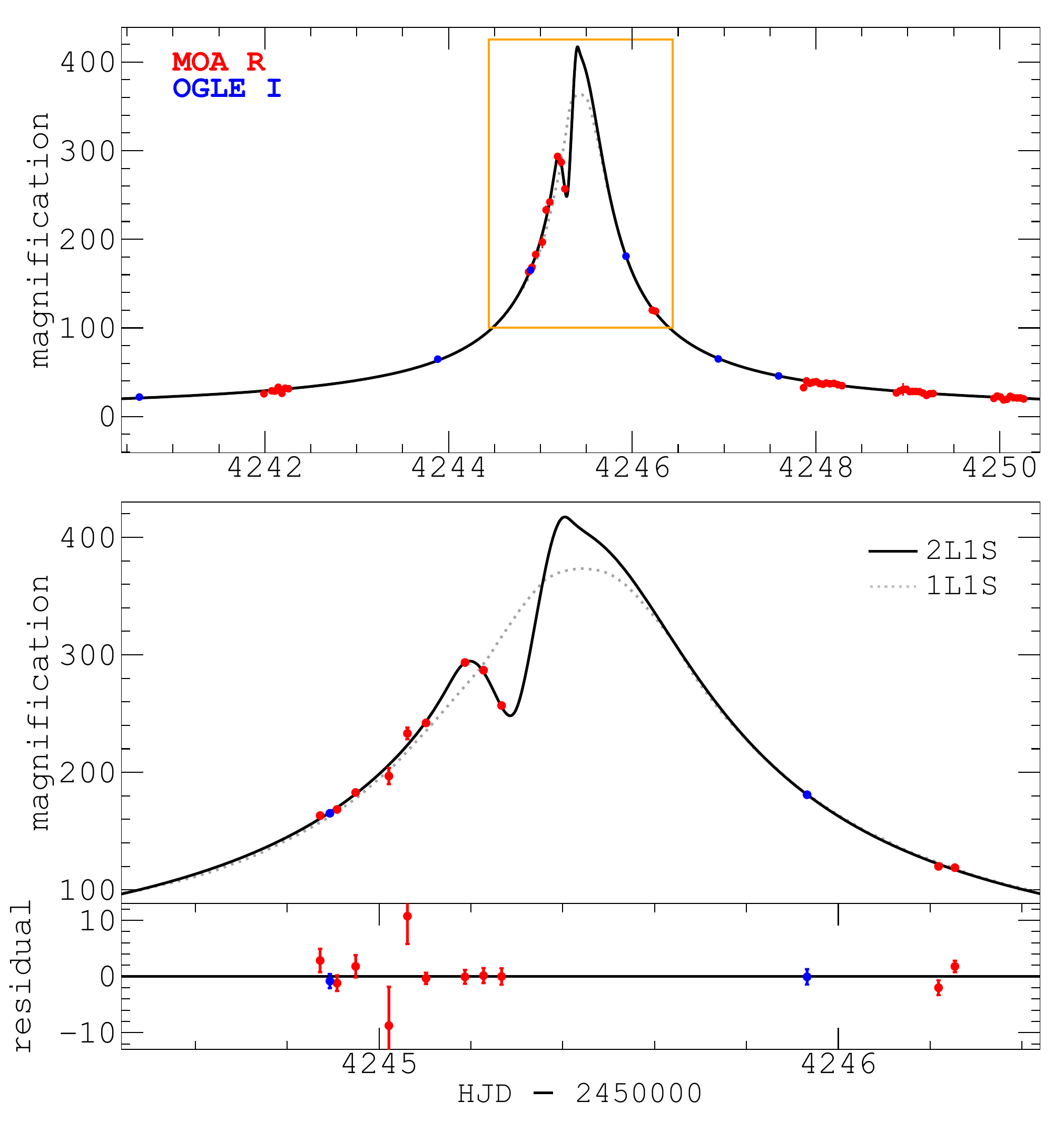}
\centering
\caption{Best fit light curve with constraints from the high resolution follow-up data as described in section \ref{sec:follow-up}. The 2L1S model shown is from the second column of Table \ref{tab:lightcurve_par} with $u_0 < 0$ and $s < 1$. The y-axis is given in flux units which are normalized to the $I_S = 21.8$ source star (e.g. `magnification') from the modeling. \label{fig:lightcurve}}
\end{figure*}

Figure \ref{fig:lightcurve} shows the observed light curve with OGLE (blue) and MOA (red) data as well as the best-fit planetary model by assuming a double-lens-single-source event (2L1S) from our re-analysis of the light curve modeling. The original light curve analysis for this event was presented by \cite{bennett:2008b} (hereafter B08). The only photometric monitoring of the target during magnification was conducted in OGLE-I and MOA-R bands. Due to the faintness of the source, there is no direct $V$-band measurement of the target from OGLE or MOA. In order to get a source color estimate, earlier studies used the photometric measurements from these two data sets and converted to (V-I) color following \cite{Gould2010}.
As apparent in Figure \ref{fig:lightcurve}, there are significant gaps in the photometric coverage for this event. Due to this incomplete coverage, there are multiple binary lens solutions, with similar mass-ratio that can equally explain the deviations in the light curve due to a binary lens system. 

This lack of coverage also resulted in large uncertainties in the measurement of the angular source size and a poorly determined angular Einstein radius, $\theta_E$. However, these various solutions all gave a low mass planetary system with a mass ratio of $q \sim 2\times10^{-4}$, and with quite large errors on the reported $q$'s. Using the constraints from microlensing parallax and the source star size, B08 concluded that the lens system was composed of a $0.06^{+0.028}_{-0.021} M_\odot$ object orbited by a $3.3^{+4.9}_{-1.6}~M_\oplus$ super-Earth. We note at the time of the B08 publication the MOA team was unaware of systematics in their photometry due to chromatic differential refraction effects \citep{bennett:2012a}. This led to an erroneous measurement of microlensing parallax ($\pi_E$) reported in their study. Further, the caustic-crossing models presented in B08 contributed to relatively small error bars on the derived planet mass (see Figure 5 in B08). These caustic-crossing models have now been largely ruled out by this study, therefore the planet mass error bars have increased (see Section \ref{sec:lens-properties}).\\

\subsection{Constraining the lensing system with adaptive optics} \label{subsec:kubas-work}

\cite{kubas:2012a} (hereafter K12) obtained two epochs with NACO AO imaging on the Very Large Telescope (VLT) shortly after the peak of the microlensing event when the target was still magnified by a factor of 1.23, as well as 18 months later at baseline. They observed in three bands, $J,~H,~K_S$, this was the first microlensing event for which a fairly large AO data set had been obtained. The AO data was reduced with the Eclipse package \citep{devillard:1997a} and the authors performed PSF photometry using the Starfinder tool \citep{diolaiti:2000a}. The absolute calibration was performed by a two-stage process using 2MASS and data collected by the IRSF telescope in South Africa. Knowing the source flux from the microlensing fit, NACO AO detected excess flux in all three near-IR bands. Assuming that all the excess flux comes from the microlensing images and the lens brightness, they obtained new constraints on the lensing system. Combining the results of the two epochs, they derived that the lens has the following magnitudes: $J_L = 20.73 \pm 0.32,~H_L = 19.94 \pm 0.35,~K_L = 19.16 \pm 0.20$. Using these constraints, and the (erroneous) microlensing parallax fit by B08, they concluded that the lensing system is a $0.084^{+0.015}_{-0.012} M_\odot$ M dwarf at a distance of 
$660^{+100}_{-70}$ pc orbited by a $3.2^{+5.2}_{-1.8}~M_\oplus$ super-Earth at $0.66^{+0.51}_{-0.22}$ AU.

\subsection{Why revisiting this system?}
MB07192 is an important event from the \citet{suzuki:2016a} sample of cold planets. Its mass ratio is in the region where a change of slope has been observed in the mass-ratio function. Also MOA have recently improved their photometry methods, so we have re-reduced the MOA photometry following \cite{bond17}. This re-reduction includes corrections for systematic errors due to chromatic differential refraction \citep{bennett:2012a}. This has direct consequences on the microlensing model compared to the initial studies, which affects the fitting parameters like microlensing parallax, finite size of the source star, and other higher order effects. Additionally, over the years we have refined our procedures to process, analyze, and calibrate AO data as well as update extinction correction calculations. We will therefore adopt our standard method described by \citep{beaulieu:2018a} and re-analyze the $K_S$ NACO data. 

\indent Finally, we have obtained recent Keck-NIRC2 and \textit{HST} observations in 2018 and 2023, which should give us the opportunity to independently resolve the source and lens and measure the magnitude and direction of their relative proper motion.

\begin{deluxetable*}{lccccccc}[!htb]
%\tabletypesize{\scriptsize}
\tablecaption{New \textit{HST} and Keck observations in this work
\label{tab:hst_keck_obs_table}}
\setlength{\tabcolsep}{11pt}
\tablewidth{\columnwidth}
\tablehead{
    \colhead{Epoch (UT)} &
    \colhead{Instrument} &
    \colhead{PA} &
    \colhead{Filter} &
    \colhead{$T_{\textrm{exp}}$} &
    \colhead{$N_{\textrm{exp}}$} &
    \colhead{$\Delta t$} &
    \colhead{Reference}\\
    \colhead{(yyyy-mm-dd)} &
    \colhead{} &
    \colhead{(deg)} &
    \colhead{} &
    \colhead{(sec)} &
    \colhead{} &
    \colhead{(yr)} &
    \colhead{}}
\startdata
2012-03-30 & WFC3$-$UVIS & 131.8 & F606W (V) & 1760 & 8 & 4.85 & (a) \\ 
 & & & F814W (I) & 1640 & 8 & & \\ 
 & {\bf{\color{purple}WFC3$-$IR}} & {\bf{\color{purple}131.8}} & {\bf{\color{purple}F125W}} (J) & {\bf{\color{purple}1412}} & {\bf{\color{purple}8}} & & \\
 & {\bf{\color{purple}WFC3$-$IR}} & {\bf{\color{purple}131.8}} & {\bf{\color{purple}F160W (H)}} & {\bf{\color{purple}1059}} & {\bf{\color{purple}8}} & & \\
 2014-03-30 & WFC3$-$UVIS & 131.8 & F606W (V) & 1760 & 8 & 6.85 & (b) \\ 
 & & & F814W (I) & 1640 & 8 & & \\ 
 2018-08-06$^*$ & NIRC2 & 0.0 & $K_S$ & 900 & 15 & 11.20 & (c) \\ 
 2023-08-06 & WFC3$-$UVIS & 309.5 & F814W (I) & 600 & 2 & 16.20 & \,\,\,(d) 
\enddata
%\centering
\tablenotetext{}{\footnotesize{\textbf{Note}: $\Delta t$ gives the amount of time (in years) since the peak of the microlensing event.\\
$^*$ The 2018 epoch is from Keck, all other epochs are from \textit{HST}.\\
\textbf{References.} (a) \cite{bennett:2012prop}, (b) \cite{bennett:2014prop}, (c) \cite{bennett_KSMS}, (d) \cite{sahu:2023prop}}}
\end{deluxetable*}

%----------------------------------------------------------High Res Followup------------------------------------------------------------------------------------------------------------

\section{High angular resolution follow-up with HST and KECK} \label{sec:follow-up}

\subsection{Preparing the absolute calibration data set}\label{sec:calib}d
We use our own re-reduction of data from the VVV survey \citep{minniti:2010a} obtained with the 4m VISTA telescope at Paranal \citep{beaulieu:2018a}. We cross identified these $JHK_S$ catalogues with the VI OGLE-III map \citep{udalski:2015a}. We then obtain an OGLE-VVV catalogue of 8500 objects with $VIJHK_S$ measurements, covering the footprint of the \textit{HST} and Keck observations. We subsequently used this catalog to calibrate the \textit{HST} and Keck data, and we also revisit the VLT/NACO data. Table \ref{tab:hst_keck_obs_table} summarizes the \textit{HST} and Keck observations that are presented for the first time in this work. The data spans the years 2012 to 2023.

\subsection{Keck NIRC2}
\label{sec:keck-followup}
The target MB07192 was observed with the NIRC2 instrument on Keck-II in the $K_\textrm{short}$ band ($\lambda_{c} = 2.146 \mu m$, hereafter Ks) on August 5 and 6, 2018. The two nights of data were combined using the KAI reduction pipeline \citep{lu:2022a}. The pipeline registers the images together, applies flat field correction, dark subtraction, as well as bad pixel and cosmic ray masking before producing the final combined image that we analyze. The data from both nights are of similar quality, with an average point spread function (PSF) full width at half maximum (FWHM) of 66.2 mas for the August 5 data, and 67.5 mas for the August 6 data.

\begin{figure*}
%\figurenum{1}
\includegraphics[width=0.7\linewidth]{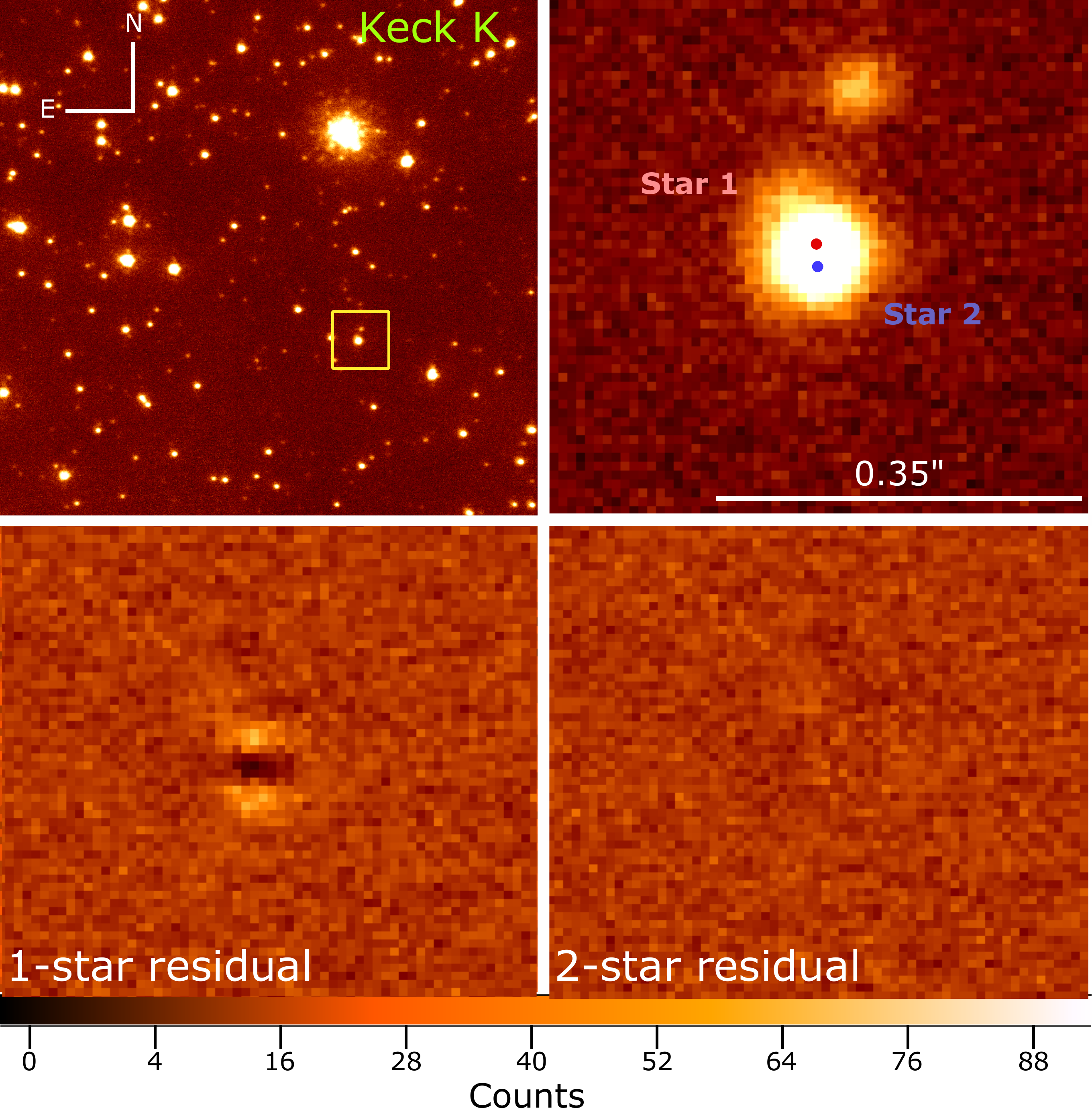}
\centering
\caption{\textit{Top Left:} The co-added sum of 15 Keck NIRC2 narrow camera images, each with an exposure time of 60 seconds. The target is indicated with a square outline. \textit{Top Right:} Zoomed image of the MB07192 blended source and lens stars. The magnitude of the separation in this epoch is $29.3 \pm 1.1$ mas. \textit{Bottom Left:} The residual image from a single-star PSF fit with DAOPHOT. A clear signal is seen due to the blended stellar profiles. \textit{Bottom Right:} The residual image for a simultaneous two-star PSF fit, showing a significantly improved subtraction. The color bar represents the pixel intensity (or counts) in the bottom panel residual images. \label{fig:quad-panel}}
\end{figure*}

\indent For the 2018 Ks band observations, both the NIRC2 wide and narrow cameras were used. The pixel scales for the wide and narrow cameras are 39.69 mas/pixel and 9.94 mas/pixel, respectively. All of the images were taken using the Keck-II laser guide star adaptive optics (LGSAO) system. For the narrow data, we combined 15 flat-field frames, six dark frames, and 15 sky frames for calibrating the science frames. A total of 15 Ks band science frames with an integration time of 60 seconds per frame were reduced using KAI which corrects instrumental aberrations and geometric distortion \citep{ghez:2008a, lu:2008a, yelda:2010a, service:2016a}. 

Because of the potentially significant effects of a spatially varying PSF in ground-based AO imaging \citep{terry:2023a}, we made a careful selection of bright and isolated reference stars that were used to build the empirical PSF model. Each of the eight selected PSF reference stars has magnitude $-0.7 < m < 0.7$ and separations $-4\arcsec < r < 4\arcsec$ from the target. The resulting PSF model has a FWHM in the x and y directions of 6.8 pixels and 6.5 pixels, respectively.

Further, a co-add of 4 wide camera images were used for photometric calibration using the catalogue prepared in section \ref{sec:calib}. The wide camera images were flat-fielded, dark current corrected, and stacked using the \texttt{SWarp} software \citep{bertin:2010a}. We performed astrometry and photometry on the co-added wide camera image using SExtractor \citep{bertin:1996a}, and subsequently calibrated the narrow camera images to the wide camera image by matching 80 bright isolated stars in the frames. The uncertainty resulting from this procedure is 0.05 magnitudes. Table \ref{table:HST_KECK} gives the calibrated magnitude for the target as measured in all high-resolution epochs detailed in this work.\\

\begin{deluxetable*}{llllll}[]
\deluxetablecaption{\textit{HST}, VLT NACO, and Keck single-star PSF photometry\label{table:HST_KECK}}
\tablecolumns{8}
\setlength{\tabcolsep}{10.0pt}
\tablewidth{\linewidth}
%\tableheight{12pt}
\tablehead{
\colhead{\hspace{-1.3cm}Data set} &
\colhead{$V$} & \colhead{$I$} & \colhead{$J$}
& \colhead{$H$}& \colhead{$K_S$} 
}
\startdata
$HST$ 2012 & $23.88\pm 0.02$ & $20.93\pm 0.01$ & $19.04\pm 0.01$ & $18.28\pm 0.01$ & \\
$HST$ 2014 & $23.83\pm 0.02$ & $20.90\pm 0.01$ & & & \\
K12 NACO ep.1 & & & $19.21 \pm 0.04$ & $18.28 \pm 0.04$ & $17.95 \pm 0.04$ \\
K12 NACO ep.2 & & & $19.32 \pm 0.07$ & $18.55 \pm 0.11$ & $17.99 \pm 0.04$ \\
NACO ep.1 & & & & & $17.80 \pm 0.05$ \\
NACO ep.2 & & & & & $17.92 \pm 0.05$ \\
Keck 2018 & & & & & $17.88 \pm 0.05$ \\
$HST$ 2023 & & $21.01\pm 0.04$ & & & \\
\enddata
\tablenotetext{}{\footnotesize{\textbf{Note}: We provide the magnitudes measured at the source position for MB07192. We recall the measured magnitudes from K12 for the two epochs. We underline that the time of the first epoch, the source was still amplified by $\sim 0.15$ mag. We re-analyzed the NACO $K_S$ images and calibrated against VVV for the two epochs. Finally, we provide our flux calibration in $K_S$ with Keck-NIRC2.}}
\end{deluxetable*}

\subsection{The Extinction Towards the Source star}\label{sec:extinction}
The OGLE extinction calculator\footnote{https://ogle.astrouw.edu.pl/cgi-ogle/getext.py} is a standard way to estimate the extinction for a galactic bulge field, and it has been commonly used for many years. The calculator is derived from the reddening and extinction study of \cite{nataf:2013a}. For MB07192, the calculator gives 
$E(V - I) = 1.10 \pm 0.127$ and an extinction $A_I = 1.3$. These standard extinction maps have recently been superseded by the \cite{Surot2020} analysis of the VVV survey and give $E(J-Ks) = 0.329 \pm 0.018$ at the location of the target. 
We then follow \cite{nataf:2013a} in adopting $E(J-K_S)/E(V-I) = 0.3433$, and $A_I=0.7465 E(V-I)+1.37 E(J_{K_S}),$ with which we derive the extinctions. Following \cite{nishiyama:2009a}, we obtain the extinctions summarized in Table \ref{table:extinction} along with prior estimates from the literature. For our subsequent analysis, we adopt the numbers from the last row of Table \ref{table_extinction} (i.e. this work).

\subsection{Resolving the Source and Lens in Keck/NIRC 2}\label{sec:keck_photometry}
Given the lens detections from \textit{HST} 2012 and 2014 data, the lens and source stars have a predicted separation of $0.65\times$FWHM in 2018, we expect the stars to be partially resolved, so it is necessary to use a PSF fitting routine to measure both targets separately. Following the methods of \cite{bhattacharya:2018a} and \cite{terry:2021a}, we use a modified version of the DAOPHOT-II package \citep{stetson:1987a}, that we call \textit{DAOPHOT\_MCMC}, to run Markov Chain Monte Carlo (MCMC) sampling on the pixel grid encompassing the blended targets. Further details of the MCMC routine are given in \cite{terry:2021a, terry:2022a}.
\\
\indent The stellar profile does not appear to be significantly extended in the NIRC2 data, as seen in the top-right panel of Figure \ref{fig:quad-panel}. However, using \textit{DAOPHOT\_MCMC} to fit a single-star PSF to the target produces the residual seen in the lower-left panel of Figure \ref{fig:quad-panel}, which shows a strong signal due to extended flux from the blended star (presumed lens). Re-running the routine in the two-star fitting mode (e.g. simultaneously fit two PSF models) produces a significantly better fit as expected, with a $\chi^2$ improvement of $\Delta\chi^2 \sim 784$. The two-star residual is nearly featureless, as can be seen in the lower-right panel of Figure \ref{fig:quad-panel}. Table \ref{table:dual-phot} shows the calibrated magnitudes for the two stars of $K_{1}=18.94 \pm 0.10$ and $K_{2}=18.39 \pm 0.09$.

\indent The final error bars on the Keck photometry and astrometry that we report in Tables \ref{table:dual-phot} and \ref{table:epoch_seps} are determined with a combination of MCMC and jackknife errors. The jackknife method \citep{quenouille49, quenouille56, tierney1999a} allows us to determine uncertainties due to PSF variations between individual Keck images. From the total of 15 Keck images, we construct $N = 14$ co-added jackknife images, with each combined image containing all but one successive image in each iteration. This method is also sometimes called the ``drop-one" or ``leave-one-out" method. The jackknife errors are then calculated via the equation:

\begin{equation}
    \sigma_x = \sqrt{\frac{N-1}{N}\sum (x_i - \bar{x})^2},
\end{equation}

\noindent where $x_i$ is a given value for the $i$th jackknife image, and $\bar{x}$ is the mean value for the jackknife images. See \cite{bhattacharya:2021a} and \cite{terry:2022a} for further details on the jackknife method.

\indent From the dual-star PSF fitting in Keck, we find a difference in $K$-band magnitude between the two blended stars of $K_{S1} - K_{S2} = -0.55 \pm 0.13$. Since the two stars are similar enough magnitude in $K$, at this point we simply apply arbitrary labels of `star 1' and `star 2' to the two stars in Keck. However, our subsequent analysis of the \textit{HST} data will allow us to confidently determine which star is the source and which is the lens (Section \ref{subsec:source_lens_mag}).

\subsection{HST WFC3/UVIS: 2012-2014-2023 Data} \label{sec:hst-followup}
The target MB07192 was observed a total of three times with the WFC3/UVIS camera on the Hubble Space Telescope (\textit{HST}). The first observation took place on 23 February 2012 in the F555W, F814W, F125W and F160W filters. A second epoch of observations were obtained on 30 March 2014 with the same four filters, and finally a third epoch was obtained on 06 August 2023 with just two exposures in the F814W filter.  The datasets are from proposals GO-12541 (PI: Bennett), GO-13417 (PI: Bennett), and GO-16716 (PI: Sahu), and were obtained from the Mikulski Archive for Space Telescopes (MAST). We flat-fielded, stacked, corrected for distortions and performed PSF photometry with the program \texttt{DOLPHOT} \citep{Dolphin2016}. Because of the disparate sensitivities, the visible images obtained with the UVIS module (F555W and F814W) and the near-IR images obtained with the IR module (F125W and F160W, Section \ref{sec:hst-IR-followup}) were reduced separately.

The drizzled, stacked frames with the astrometric solutions from the Space Telescope Science Institute (STScI) were used as the reference image for source finding. We used \texttt{DOLPHOT} to correct for pixel area distortions, remove cosmic rays, and perform PSF fitting photometry of the individual frames. Because of the crowded nature of the field, the sky background was determined iteratively and many artifacts due to bright stars were rejected. \texttt{DOLPHOT} uses a library of reference PSFs for each filter and applies a perturbation to the $N\times N$ array of PSFs based on differences between the PSF present in the image and the library PSF \citep{anderson:2006a}. This perturbation typically adjusts the central pixels of the PSF by a few percent, where this difference comes mostly from telescope breathing or jitter. The PSF-fit magnitudes are corrected to a standard circular aperture of radius 0$\farcs$5 and matched across all filters. In order to eliminate marginal detections and stars badly impacted by bright neighbors, we rejected stars with signal-to-noise ratio $<$5 and crowding parameter $>$0.75 as well as any objects flagged by the software as too sharp or too extended to be stellar.  The output magnitudes are given in the STScI VegaMag system (m555, m814, m125, m160). Note that we used only main sequence stars for calibration, and ignore color terms between VVV and the STScI VegaMag system.\\

\subsection{HST WFC3/IR: 2012 Data} \label{sec:hst-IR-followup}
 For the 2012 \textit{HST} epoch, the WFC3-IR channel was utilized to take eight exposures with the $F125W$ ($\lambda_c = 1.248\, \mu m$) filter and eight exposures with the $F160W$ ($\lambda_c = 1.537\, \mu m$) filter. These are wide $J$ and $H$-band filters, respectively. Similar to the reduction procedure described in the previous Section \ref{sec:hst-followup}, \texttt{DOLPHOT} was used for flat-fielding, distortion corrections, pixel area map corrections, cosmic ray rejection, and PSF-fitting which gives the resulting photometry and astrometry for all detected sources in the field. \\
\indent In contrast to the WFC3-UVIS and Keck/NIRC2 data, the source and lens were not independently resolved in the WFC3-IR data. This is primarily due to the much larger pixel size in near-IR \textit{HST} images (${\sim}100$ mas/pix), and the fact that the only WFC3-IR data was taken in the earliest \textit{HST} epoch (2012) when the lens and source were more highly blended than they were in the 2014 or 2023 \textit{HST} epochs. It's likely the lens and source may have been at least partially resolved if near-IR data was taken in 2014, and very likely in 2023. Details of the 2012 WFC3-IR visit can be found in Table \ref{tab:hst_keck_obs_table}, and the single-star PSF photometry for the target (source $+$ lens) can be found in columns 4 and 5 of Table \ref{table:HST_KECK}. \\
\indent Lastly, given the calibrated $J$ and $H$-band magnitudes we measured for the combined source $+$ lens in the WFC3-IR images, we measured the excess flux at the position of the source in these passbands to estimate the lens (e.g. blend) star magnitude. This assumes all of the blended light comes from the lens, but we can be confident that this is the case since we have multi-epoch direct lens detections in the other \textit{HST} and Keck datasets. The Appendix includes Figure \ref{fig:CMD_IR} which shows the color-magnitude diagram (CMD) for all of the detected sources in the \textit{HST} field, as well as the estimated $J$ and $H$-band magnitudes and colors for the source and lens.

\subsection{\textit{HST} Multiple Star PSF Fitting} \label{sec:hst_multistar_fits}
In addition to the photometry obtained using \texttt{DOLPHOT}, we performed multi-star PSF fitting on the target in all three of the \textit{HST} epochs. Since the 2012 and 2014 epochs are approximately 6.4 and 4.4 years before the Keck observations, we expect the separation between the source and lens star to be $0.619\times$ and $0.734\times$ smaller in these \textit{HST} images compared to Keck. This is due primarily to the relative proper motion between the source and lens star as observed from Earth (and \textit{HST}). Similarly, the 2023 \textit{HST} images were taken approximately 5.0 years after the Keck observations, so we expect the lens and source separation to be $1.445\times$ larger in this epoch than the Keck images. Because each \textit{HST} observation is separated by at least several years and some epochs were taken at different position angles (PA), we performed coordinate transformations between the Keck observation and each of the \textit{HST} observations independently. We do this by cross-matching ${\sim}$two dozen isolated and bright (but not saturated) stars in each dataset, and then calculate the linear (i.e. first order) transformation between the pixel positions in the \textit{HST} and Keck catalogs. The transformations are listed as follows:

\begin{equation*}
\begin{split}
    \textrm{2012:}\hspace{1cm} x_{hst} = -0.170x_{keck} + 0.186y_{keck} + 804.011 \\
    y_{hst} = -0.184x_{keck} - 0.169y_{keck} + 1175.664 \\
    \\
    \textrm{2014:}\hspace{1cm} x_{hst} = -0.169x_{keck} + 0.186y_{keck} + 802.920 \\
    y_{hst} = -0.185x_{keck} - 0.169y_{keck} + 1175.561 \\
    \\
    \textrm{2023:}\hspace{1cm} x_{hst} = 0.164x_{keck} - 0.188y_{keck} + 283.268 \\
    y_{hst} = 0.187x_{keck} + 0.166y_{keck} + 225.342
\end{split}
\end{equation*}

\noindent The average Root mean square (RMS) scatter for these relations is $\sigma_x \sim 0.25$ and $\sigma_y \sim 0.20$ \textit{HST/UVIS} pixels for the same 16 stars used in each transformation. Given the varying baseline between the earliest and latest \textit{HST} epochs and the 2018 Keck epoch, this scatter of ${\sim}13$ mas can be at least mostly explained by an average proper motion of $\sim$2.5 mas/yr in each direction. We note the 2012 and 2014 data were taken with the larger sub-array chip, UVIS2-2K2C-SUB, while the recent 2023 data were taken with the smaller chip, UVIS2-C1K1C-SUB. Using the smaller sub-array chip allows us to minimize the negative effect of charge-transfer efficiency (CTE), since the detector has degraded between the 2012/2014 and 2023 epochs.

\begin{figure*}
%\figurenum{1}
\includegraphics[width=0.7\linewidth]{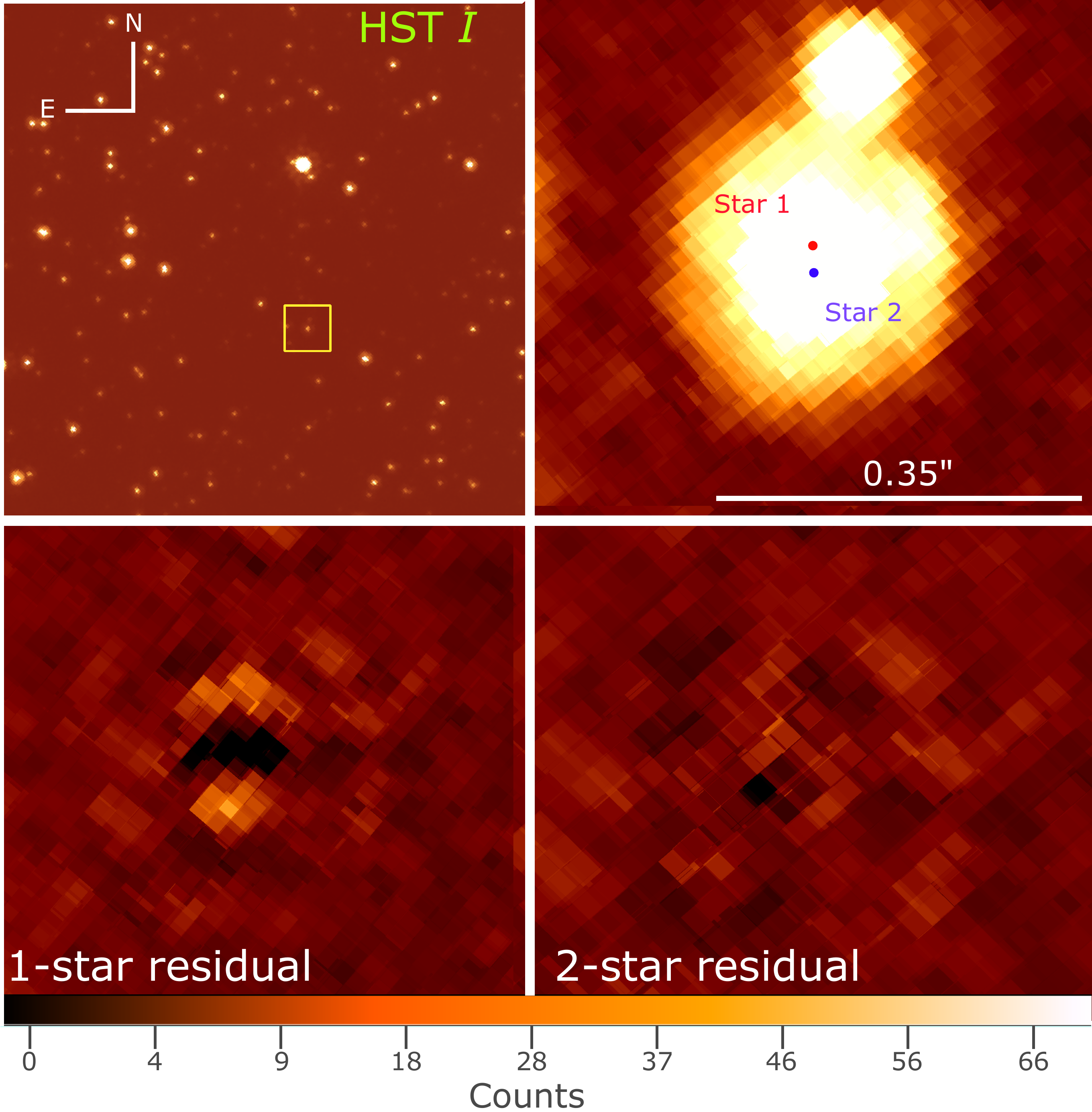}
\centering
\caption{Similar to Figure \ref{fig:quad-panel}, but for the 2014 \textit{HST} data (eight exposures). The zoomed inset and residual image panels show 100 $\times$ 100 super-sampled pixels where the observed dither offsets are accurate to 0.01 pixels. The color bar represents the pixel intensity (counts) seen in the top-right and lower-left/right panels. \label{fig:quad-panel_hst}}
\end{figure*}

\indent This \textit{HST} analysis was performed using a modified version of the codes developed in \cite{bennett:2015a} and \cite{bhattacharya:2018a}, which analyzes the original individual images with no resampling. This avoids any loss in resolution that can occur when dithered, undersampled images are combined. The top-left panel of Figure \ref{fig:quad-panel_hst} shows the target and surrounding \textit{HST} stars from the combined I-band image in 2014. A zoom on the target is shown in the top-right panel, which also shows an unrelated star to the North of MB07192. The lower panels of Figure \ref{fig:quad-panel_hst} show the residual images after fitting a single PSF model and simultaneously fitting two PSF models to the blended stars. The single-star residual shows the typical signal that we would expect for two highly blended stars. The direction and amplitude of the measured separation here is consistent with the 2018 detection in Keck (Table \ref{table:epoch_seps}). The $V$-band detection is at a lower confidence than the $I$-band detection (${\sim}2\sigma$ vs. $>5\sigma$ above the noise level). This leads to a larger error on the measured $V$-band lens magnitude (Table \ref{table:dual-phot}) and significantly larger error on the measured lens-source separations in \textit{HST} $V$-band (see Table \ref{table:epoch_seps}).

\indent Given the strong detection in the Keck data, we impose separation constraints when analyzing the earlier \textit{HST} epochs, particularly the 2012 epoch in $V$-band, where the lens detection is most marginal. We convert the Keck relative proper motion value ($\mu_\textrm{rel,H} = 2.63 \pm 0.13$ mas/yr) to constraints on the position of the lens and source in the 2012 \textit{HST} images, while taking into account the 4.8520 years between the microlensing event peak and the 2012 \textit{Hubble} observations. We note that in all of our \textit{HST} PSF fitting procedures we include the unrelated faint nearby neighbor as a third star to avoid any interference of its PSF with our measurement of lens-source separation. Between the 2012 and 2023 epochs, the unrelated neighbor star moves ${\sim}1$ \textit{HST} pixel closer to MB07192.

\indent For all three \textit{HST} epochs (2012, 2014, 2023), the F814W fits converge to a consistent solution with `star 1', to the North as the slightly brighter star ($\Delta m_{\textrm{F814W}}\, {\sim}\, $0.1). For the two epochs of F555W data (2012, 2014), the PSF fit converged to a unique solution in the 2014 data without requiring any separation constraint but the 2012 fit required a separation constraint to be imposed, as mentioned previously. In all \textit{HST} F814W fits, `star 2', to the South is slightly fainter than `star 1'. Our reduction and fitting code places the star coordinates from both filters into the same reference system, so all stars have positions that are consistent between both passbands. The best-fit magnitudes (calibrated to OGLE $V$ and $I$) from the 2014 \textit{HST} epoch are given in Table \ref{table:dual-phot}, and the best-fit positions in all epochs and filters are given in Table \ref{table:epoch_seps}.

\indent The \textit{HST} data were calibrated to the OGLE-III catalog \citep{szymanski:2011a} using eight relatively bright isolated OGLE-III stars that were matched to \textit{HST} stars. The same eight stars were used in each epoch. For the best quality \textit{HST} data in both filters (i.e. 2014 epoch), the calibrations yielded $I_1 = 21.56 \pm 0.15$, $V_1 = 24.93 \pm 0.32$, $I_2 = 21.68 \pm 0.16$, and $V_2 = 24.25 \pm 0.18$. The magnitude of both lens and source stars combined is measured to significantly higher precision, $I_{12} = 20.87 \pm 0.02$ and $V_{12} = 23.79 \pm 0.04$. This combined magnitude allows us to place a stronger constraint when re-evaluating the light curve photometry. During our PSF fitting, the two blended stars can trade flux back and forth which results in larger errors on the individual stars' magnitude.

\begin{deluxetable*}{llllllll}[]
\deluxetablecaption{Extinction estimates towards the source\label{table:extinction}}
\tablecolumns{8}
\setlength{\tabcolsep}{4.0pt}
\tablewidth{\linewidth}
%\tableheight{12pt}
\tablehead{
\colhead{\hspace{-0.4cm}Ext. map} &
\colhead{$E(V-I)$} & \colhead{$E(J-K_s$)} & \colhead{$A_V$}
& \colhead{$A_I$}& \colhead{$A_J$}& \colhead{$A_H$}& \colhead{$A_{K_S}$}
}
\startdata
B08 & $1.12 \pm 0.09$ &  & $2.73 \pm 0.13$ & 
    $1.61 \pm 0.10$& & & \\
K12 &    & $0.43 \pm 0.14$&  & & 
$0.72 \pm 0.10$ & $0.46 \pm 0.10$   & $0.29 \pm 0.10$\\
ext. calc & $1.10 \pm 0.13$ &  & $2.43 \pm 0.16$ & 
    $1.33 \pm 0.1$ & & & \\\
\hspace{-0.25cm} {\bf this study }&  $\bf 1.10 \pm 0.06$ &  $\bf 0.33 \pm 0.02$ & $\bf 2.45 \pm 0.15$ & 
   $\bf 1.35 \pm 0.07$ & $\bf 0.44 \pm 0.02$ & $\bf 0.24 \pm 0.01$ & $\bf 0.11 \pm 0.01$\\
\enddata
\tablenotetext{}{\footnotesize{\textbf{Note}. We summarize here the different
estimates for the extinction towards the source, in the initial study (B08), the follow up work with NACO data (K12), and this study. Extinction values are derived from a combination of the methods described in \cite{nishiyama:2009a}, \cite{bennett:2010b}, \cite{nataf:2013a}, and \cite{Surot2020} (see section \ref{sec:extinction}).}}
\label{table_extinction}
\end{deluxetable*}

\subsection{Identifying the Source and Lens Stars} \label{subsec:source_lens_mag}
\begin{figure*}
\centering
\includegraphics[width=0.75\linewidth]{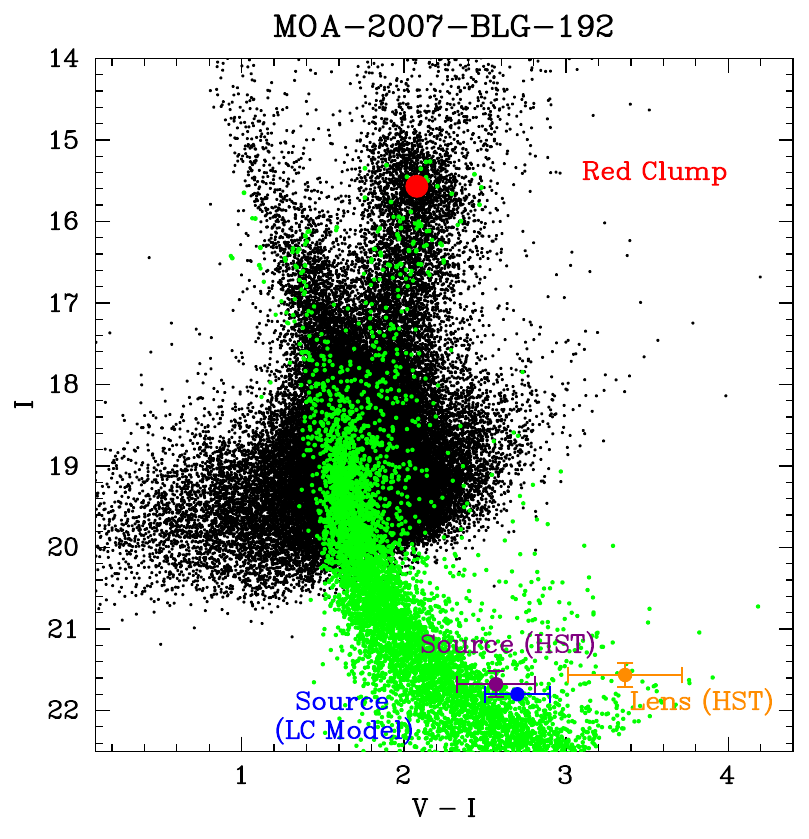}

\centering
\caption{The observed color-magnitude diagram (CMD) for the MB07192 field. The OGLE-III stars within 90 arcseconds of MB07192 are shown in black, with the \textit{HST} CMD of all detected sources from the 2014 epoch shown in green. The red point indicates the location of the red clump centroid, and the purple and orange points show the source and lens colors and magnitudes from the 2014 \textit{HST} observations.  The blue point indicates the source star magnitude and color given by the original light curve modeling.
\label{fig:CMD}}
\end{figure*}

With the \textit{HST} $V$ and $I$-band measurements described in Section \ref{sec:hst_multistar_fits}, we can now attempt to determine which star is the source and which is the lens. As mentioned previously, since the original discovery paper of \cite{bennett:2008a}, the MOA group has begun detrending its photometry to remove systematic errors caused by differential atmospheric refraction \citep{bennett:2002a, bond17}. Following \cite{bond17}, we correct the MOA photometric data and perform re-modeling of the MOA + OGLE photometry. This re-analysis yields an estimate of the source star $I$-band magnitude of $I_S = 21.8 \pm 0.05$ with a color of $V_S - I_S = 2.7 \pm 0.2$. This source $I$-band magnitude is within $1\sigma$ of the \textit{HST} $I$-band magnitude for `star 2', and just over $1\sigma$ fainter than the \textit{HST} $I$-band magnitude for `star 1'. Additionally, this estimated source color is a closer match to the measured \textit{HST} $V - I$ color of `star 2' ($2.57 \pm 0.24$) as can be seen in Figure \ref{fig:CMD}. These results support the identification of `star 2' as the true source star. However, since the ground-based $V$-band estimate of the source comes from a relatively weak relationship (OGLE I $-$ MOA R), we conduct a further verification of the source and lens using their relative proper motions as measured in \textit{HST} and Keck.

\begin{figure*}
\centering
\includegraphics[width=0.75\linewidth]{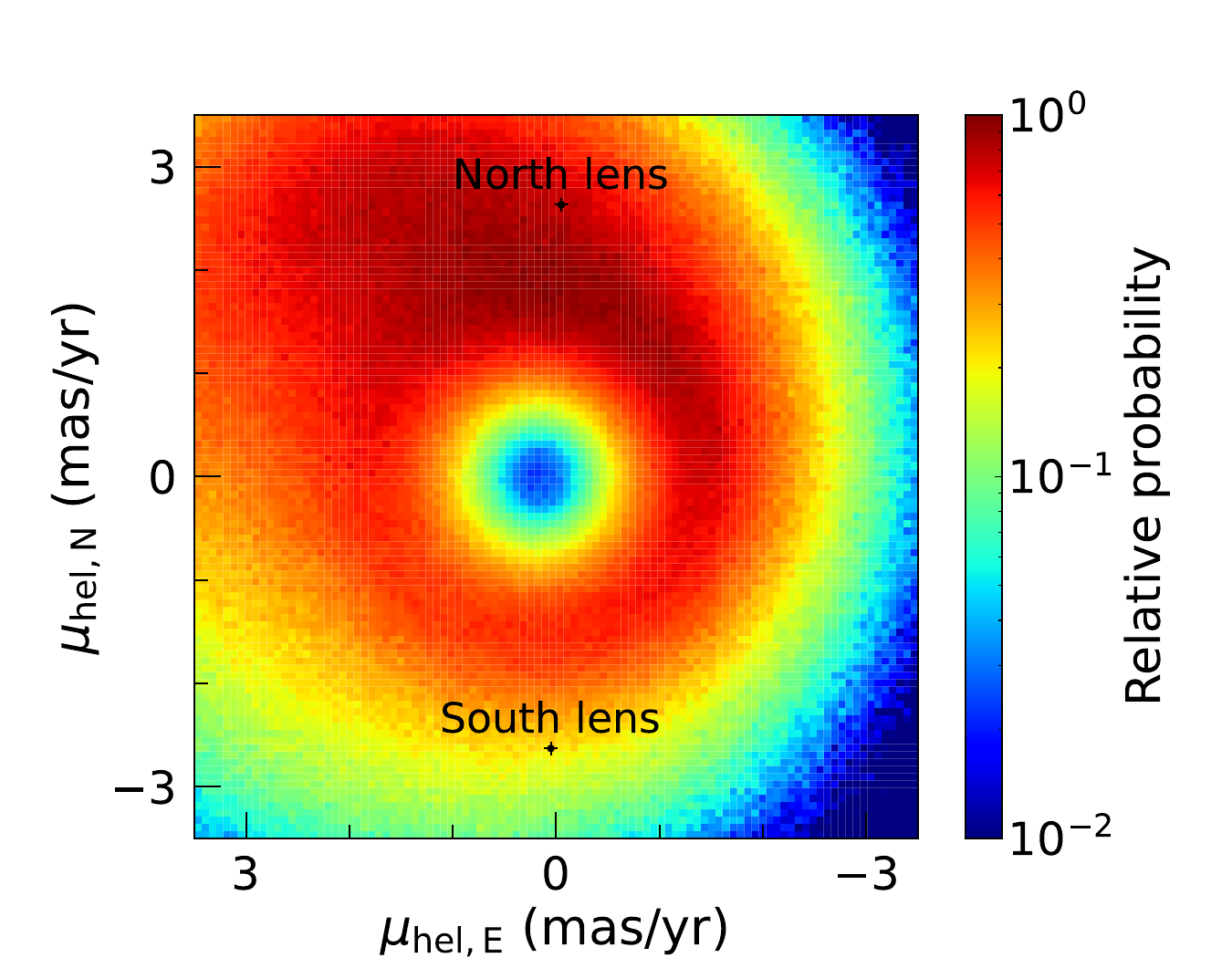}

\centering
\caption{The probability distribution for the north and east components of lens-source relative proper motion ($\mu_\textrm{rel}$) using the Galactic model from \cite{koshimoto:2021a} and \texttt{genulens} \citep{koshimoto:code}. The possible lens positions (North and South) are plotted in black and given by the relative motion of the two stars detected in the \textit{HST} and Keck data. Importantly, this distribution uses $t_E$ values that are close to the measured $t_E$ value from the light curve modeling ($t_E\, {\sim}\, 99.5$ days). This implies that the North star is $>2\times$ more likely to be the lens than the South star. \label{fig:murel_distribution}}
\end{figure*}

\indent We calculate the 2D prior probability distribution of the lens-source relative proper motion ($\mu_{\textrm{rel}}$) using the \cite{koshimoto:2021a} Galactic model to determine which stars are the preferred lens and source. Figure \ref{fig:murel_distribution} shows this proper motion distribution for MB07192, with two locations for the possible lens (the `North' or `South' star). We calculate these $\mu_{\textrm{rel}}$ priors from the distribution of single lens stars that reproduces the Einstein radius crossing time that accounts for the host star mass in a binary lens case, i.e., $t_E/\sqrt{1 + q}$. The results show that there is a preference for the `North star' to be the true lens star considering the stellar distribution along this sight-line. The relative probability is $P_N / P_S = 25.88/12.43 = 2.08$; this means the `North star' is $>2\times$ more likely to be the lens than the `South star'. So, given the locations of `Star 2' and `Star 1' on the CMD (before re-labeling them) and the relative proper motion prior probability distribution (Figure \ref{fig:murel_distribution}), we identify `Star 2' (e.g. the `South star') to be the true source star and `Star 1' (e.g. the `North star') to be the true lens star which hosts the planet. We subsequently label the source and lens on the CMD in Figure \ref{fig:CMD} as well as the stars in Table \ref{table:dual-phot}.

%%%%%%%%%%%
%%%%%%%%%%%

\begin{deluxetable*}{llll}[!htb]
\deluxetablecaption{\textit{HST} and Keck multi-star PSF photometry\label{table:dual-phot}}
\tablecolumns{4}
\setlength{\tabcolsep}{18.0pt}
\tablewidth{\linewidth}
%\tableheight{12pt}
\tablehead{
\colhead{\hspace{-1.7cm}Star} &
\colhead{$V$ Mag} & \colhead{$I$ Mag} & \colhead{$K$ Mag}
}
\startdata
Star 1 (Lens) & $24.93 \pm 0.32$ & $21.56 \pm 0.15$ & $18.39 \pm 0.09$\\
Star 2 (source) & $24.25 \pm 0.18$ & $21.68 \pm 0.16$ & $18.94 \pm 0.10$\\
Lens $+$ Source & $23.79 \pm 0.04$ & $20.87 \pm 0.02$ & $17.88 \pm 0.05$\\
\enddata
\tablenotetext{}{\footnotesize{\textbf{Note}. $V$ and $I$ magnitudes are calibrated to the OGLE-III system and $K$ magnitudes are calibrated to the 2MASS system as described in section \ref{sec:follow-up}.}}
\end{deluxetable*}

%----------------------------------------------------------Proper Motion------------------------------------------------------------------------------------------------------------

\section{Lens-Source Relative Proper Motion} \label{sec:prop-motion}

The Keck (2018) and HST (2012, 2014, 2023) follow up observations were taken between 4.85 and 16.20 years after the peak magnification which occurred in May 2007. The motion of the source and lens on the sky is the primary cause for their apparent separation, however there is also a small component that can be attributed to the orbital motion of Earth (e.g. trigonometric parallax). As this effect is of order $\leq0.10$ mas for a lens at a distance of $D_{L} \geq 2$ kpc, we are safe to ignore this contribution in our analysis as it is much smaller than the error bars on the stellar position measurements  (e.g. the astrometric measurements given in Table \ref{table:epoch_seps}). The mean lens-source relative proper motion is measured to be $\mu_{\textrm{rel},H} = (\mu_{\textrm{rel,H,E}},\mu_{\textrm{rel,H,N}}) = (0.634 \pm 0.291, 2.761 \pm 0.274)$ mas yr$^{-1}$, where `H' indicates that these measurements were made in the heliocentric reference frame, and the `E' and `N' subscripts represent the East and North on-sky directions respectively.

\indent Our light curve modeling is performed in the geocentric reference frame that moves with the Earth at the time of the event peak.
Thus, we must convert between the geocentric and heliocentric frames by using the relation given by \cite{dong:2009b}:

\begin{equation}\label{eq:mu-rel}
\mu_{\textrm{rel,H}} = \mu_{\textrm{rel,G}} + \frac{{\nu_{\Earth}}{\pi_{\textrm{rel}}}}{AU} \ ,
\end{equation}

\noindent where $\nu_{\Earth}$ is Earth's projected velocity relative to the Sun at the time of peak magnification. For MB07192 this value is $\nu_{\Earth \textrm{E,N}} = (25.772, 1.237)$ km/sec = $(5.433, 0.261)$ AU yr$^{-1}$ at HJD$' = 4245.45$, where HJD$' = \textrm{HJD} - 2450000$. With this information and the relative parallax relation $\pi_{\rm{rel}} \equiv AU(1/D_{L} - 1/D_{S})$, we can express Equation \ref{eq:mu-rel} in a more convenient form:

\begin{equation}
    \mu_{\textrm{rel,G}} = \mu_{\textrm{rel,H}} - (5.433, 0.261) \times (1/D_{L} - 1/D_{S})\, \textrm{mas/yr},
\end{equation}

\noindent where $D_{L}$ and $D_{S}$ are the lens and source distance, respectively, given in kpc. We have directly measured $\mu_{\textrm{rel,H}}$ from the \textit{HST} and Keck data, so this gives us the relative proper motion in the geocentric frame of $\mu_{\textrm{rel,G}} = 3.10 \pm 0.19\,$mas/yr. As a reminder, the lens and source distance we use in Equation \ref{eq:mu-rel} are inferred by the best-fit light curve results which include constraints from the high-resolution imaging.

\begin{figure*}
\centering
\includegraphics[width=0.75\linewidth]{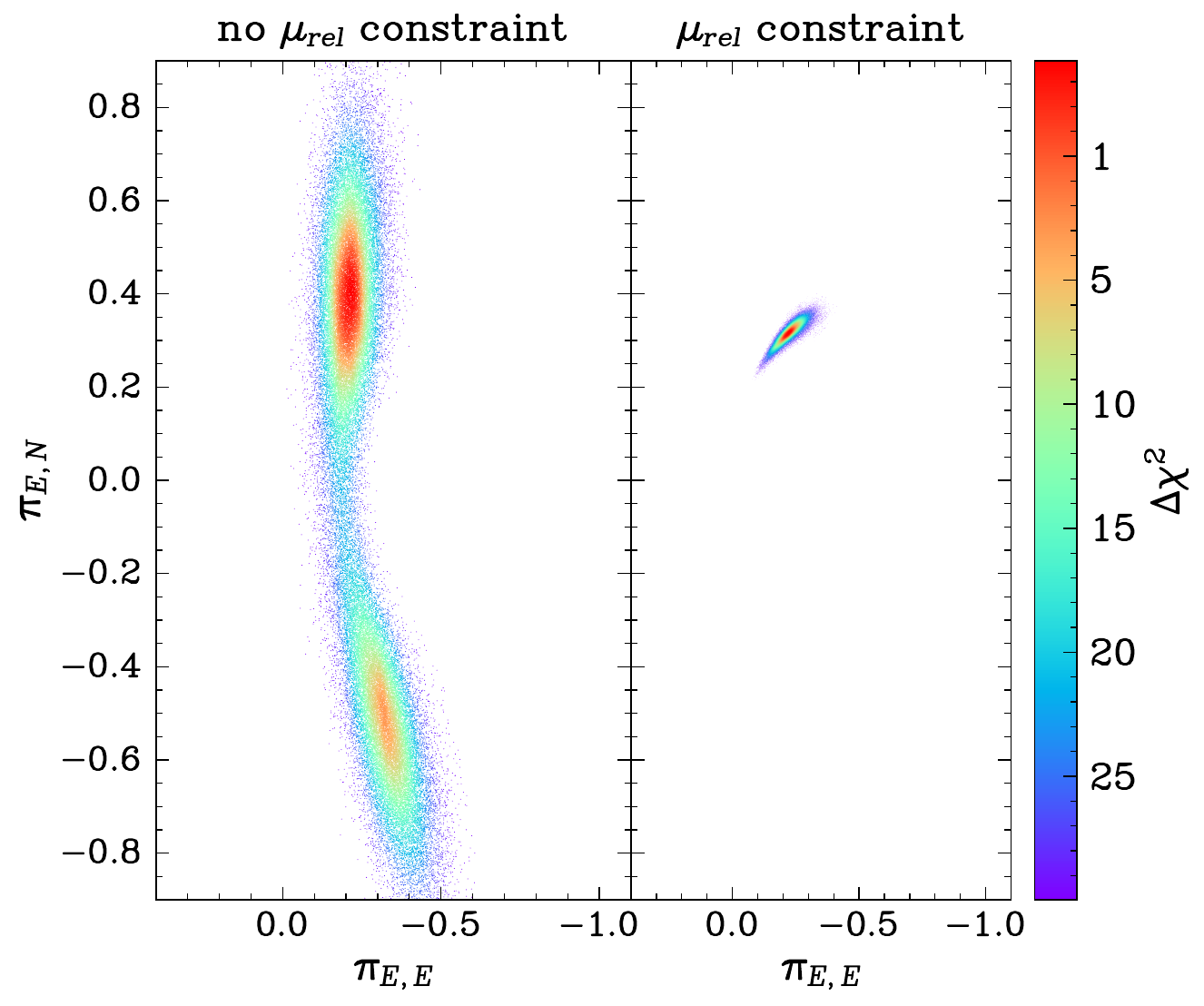}

\centering
\caption{\textit{Left}: The MCMC distribution for $\pi_E$ from the light curve modeling without any constraint from the high-resolution imaging. \textit{Right}: The MCMC distribution for $\pi_E$ from the light curve modeling with the inclusion of high-resolution imaging constraints. The color bar represents the $\chi^2$ differences from the best-fit light curve model. The two components of the relative proper motion that were measured by \textit{HST} and Keck allow the north component, $\pi_{E,N}$ to be tightly constrained.
\label{fig:piE_distribution}}
\end{figure*}

\begin{deluxetable*}{rccc}[!htb]
\deluxetablecaption{Measured Lens-Source Separations from \textit{HST} and Keck\label{table:epoch_seps}}
\tablecolumns{3}
\setlength{\tabcolsep}{14.0pt}
\tablewidth{\linewidth}
%\tableheight{12pt}
\tablehead{
\colhead{}&
  \multicolumn{3}{c}{Separation (mas)} \\
\cline{2-4}
\colhead{Year} &
\colhead{East} & \colhead{North} & \colhead{Total}
}
\startdata
$2012\, (HST\,\, V)$ & $2.28 \pm 4.60$ & $15.58 \pm 4.96$ & $15.75 \pm 6.78$\\
$(HST\,\, I)$ & $1.01 \pm 1.39$ & $18.17 \pm 1.71$ & $18.20 \pm 2.23$\\
$2014\, (HST\,\, V)$ & $9.84 \pm 4.74$ & $22.17 \pm 4.01$ & $24.26 \pm 6.22$\\
$(HST\,\, I)$ & $3.12 \pm 1.23$ & $21.62 \pm 1.03$ & $21.84 \pm 1.63$\\
$2018\, (Keck\,\, K)$ & $-0.34 \pm 1.03$ & $29.37 \pm 1.01$ & $29.38 \pm 1.46$\\
$2023\, (HST\,\, I)$ & $-1.97 \pm 1.49$ & $43.13 \pm 1.68$ & $43.17 \pm 2.26$\\
\hline
\hline
 & $\mu_{\textrm{rel,H,E}}$(mas/yr) & $\mu_{\textrm{rel,H,N}}$(mas/yr) & $\mu_{\textrm{rel,H}}$(mas/yr)\\
 \hline
weighted mean & $0.63 \pm 0.29$ & $2.76 \pm 0.27$ & $2.83 \pm 0.37$\\
\enddata
%\tablenotetext{}{\footnotesize{\textbf{Note}. Magnitudes are calibrated to the VVV system, as described in section \ref{sec:follow-up}.}}
\end{deluxetable*}

%----------------------------------------------------------Lens System Properties------------------------------------------------------------------------------------------------------------

\section{Lens System Properties} \label{sec:lens-properties}

As has been shown in prior work \citep{bhattacharya:2018a, bennett:2020a, terry:2021a, rektsini:2024a}, we find it particularly useful to apply constraints from the high resolution follow-up observations to the light curve models (we deem this ``image-constrained modeling"). This can help prevent the light curve modeling from exploring areas in the parameter space that are excluded by the high resolution follow-up observations. We refer the reader to \cite{bennett:2023a} for a full description of the methodology for applying these constraints to the modeling and an exhaustive list of the light curve + high-res imaging parameters that are important for obtaining full solutions for planetary lens systems in this context.

\indent We use the python package \texttt{eesunhong} for the light curve modeling to incorporate constraints on the brightness and separation of the lens and source stars from the high resolution imaging via \textit{HST} and Keck \citep{bennett:1996a, bennett:2010a, bennett:2023a}. Ideally, we want to use a mass-distance relation coupled with empirical mass-luminosity relations to infer the mass and distance of the host star. In order to do this, we need to know the distance to the source star, $D_S$. Thus we are required to include the source distance as a fitting parameter in the re-modeling of the light curve with imaging constraints. We include a weighting from the \cite{koshimoto:2021a} Galactic model as a prior for $D_S$, and we also use the same Galactic model to obtain a prior on the lens distance for a given value of $D_S$. This prior is not used directly in the light curve modeling, but instead is used to weight the entries in a sum of Markov chain values.

\indent The angular Einstein radius, $\theta_E$, and the microlensing parallax vector, $\bm{\pi_E}$, give relations that connect the lens system mass to the source and lens distances, $D_S$ and $D_L$ (\cite{bennett:2008b}, \cite{gaudi:2012a}). The relations are given by:

\begin{equation} \label{eq:md_thetaE}
    M_{L} = \frac{c^2}{4G}\theta_{E}^{2}\frac{D_{S}D_{L}}{D_{S}-D_{L}},
\end{equation}

\noindent and 

\begin{equation} \label{eq:md_piE}
    M_{L} = \frac{c^2}{4G}\frac{\textrm{AU}}{\bm{\pi_E}^2}\frac{D_{S}-D_{L}}{D_{S}D_{L}},
\end{equation}

\noindent where $M_{L}$ is the lens mass, $G$ and $c$ are the gravitational constant and speed of light. As mentioned previously, the measurement of $\mu_{\textrm{rel,H}}$ from the high resolution imaging allows us to measure $\mu_{\textrm{rel,G}}$ to high precision, which ultimately lets us determine $\bm{ \theta_E \sim \mu_{rel,G} \times t_E}$. Additionally, the two components of the $\mu_{\textrm{rel}}$ measurement enables a much tighter constraint on the possible values of $\pi_{E,N}$. The north direction in particular is usually only weakly constrained because it is typically perpendicular to the orbital acceleration of the observer for microlensing events towards the Galactic bulge. The geocentric relative proper motion and the microlensing parallax are related by:

\begin{equation} \label{eq:pi_E}
    \boldsymbol{\pi}_E = \frac{\pi_{\textrm{rel}}}{t_E}\frac{\boldsymbol{\mu}_{\textrm{rel,G}}}{|\mu_\textrm{rel,G}|^2},
\end{equation}

\noindent so with the measurement of $\pi_{E,E}$ and $\mu_{\textrm{rel,H}}$, we can use equations \ref{eq:mu-rel} and \ref{eq:pi_E} to solve for $\pi_{E,N}$. This tight constraint on the north component of the microlensing parallax can be seen in Figure \ref{fig:piE_distribution}, where the left panel shows the distribution in $\pi_{E,N}$ is largely unconstrained. When the constraint from the high resolution measurement of $\mu_{\textrm{rel}}$ is applied, the distribution collapses to a relatively small region centered on $\pi_{E,N}\, {\sim}\, 0.3$.

%%%%%%%%%%%%
%%%%%%%%%%%%
\begin{deluxetable*}{@{\extracolsep{4pt}}lccccc}
\deluxetablecaption{Best Fit Model Parameters with $\mu_{\textrm{rel}}$ and Magnitude Constraints \label{tab:lightcurve_par}}
\setlength{\tabcolsep}{12.0pt}
\tablewidth{\columnwidth}
\tablehead
{
\colhead{}&
  \multicolumn{2}{c}{$u_0 < 0$}&
  \multicolumn{2}{c}{$u_0 > 0$} \\
\cline{2-3} \cline{4-5}
\colhead{\hspace{-0.5cm}Parameter} & \colhead{$s < 1$}& 
\colhead{$s > 1$} & \colhead{$s < 1$} & \colhead{$s > 1$}
& \colhead{MCMC Averages}
}
\startdata
$t_E$ (days) & $99.469$ & $98.722$ & $100.111$ & $99.262$ & $99.577 \pm 3.919$\\
$t_{0}$ (HJD$'$) & $4245.446$ & $4245.448$ & $4245.431$ & $4245.436$ & $4245.440 \pm 0.0070$\\
$u_0$ & $-0.0027$ & $-0.0029$ & $0.0035$ & $0.0004$ & $-0.0027 \pm 0.0012$\\
{} & {} & {} & {} & \hspace{-2.8cm}($u_0 > 0$) & $0.00195 \pm 0.00155$\\
$s$ & $0.9102$ & $1.0311$ & $0.8780$ & $1.1441$ & $0.8728 \pm 0.0667$\\
{} & {} & {} & {} & \hspace{-2.8cm}($s > 1$) & $1.0951 \pm 0.0938$\\
$\alpha$ (rad) & $2.1061$ & $1.9288$ & $4.5473$ & $3.2862$ & $2.4364 \pm 0.5075$\\
{} & {} & {} & {} & \hspace{-2.8cm}($u_0 > 0$) & $3.9167 \pm 0.6305$\\
%$q \times 10^{-4}$ & $1.0589$ & $1.0058$ & $2.5023$ & $1.6081$ & $1.3522 \pm 1.4563$\\
$log(q)$ & $-3.9751$ & $-3.9975$ & $-3.6017$ & $-3.7937$ & $-3.8690 \pm 0.5253$\\
$t_*{}$ (days) & $0.0562$ & $0.0539$ & $0.0567$ & $0.0547$ & $0.0551 \pm 0.0044$\\
$\pi_{\textrm{E,N}}$ & $0.3161$ & $0.3152$ & $0.3119$ & $0.3133$ & $0.3154 \pm 0.0218$\\
$\pi_{\textrm{E,E}}$ & $-0.2364$ & $-0.2308$ & $-0.2338$ & $-0.2300$ & $-0.2359 \pm 0.0474$\\
$D_{\textrm{s}}$ (kpc) & $7.8423$ & $7.1562$ & $6.9687$ & $7.1156$ & $7.049 \pm 1.163$\\
Fit $\chi^2$ & $4760.94$ & $4760.97$ & $4761.45$ & $4761.53$\\
\enddata
\end{deluxetable*}

\begin{figure*}
\centering
\includegraphics[width=0.75\linewidth]{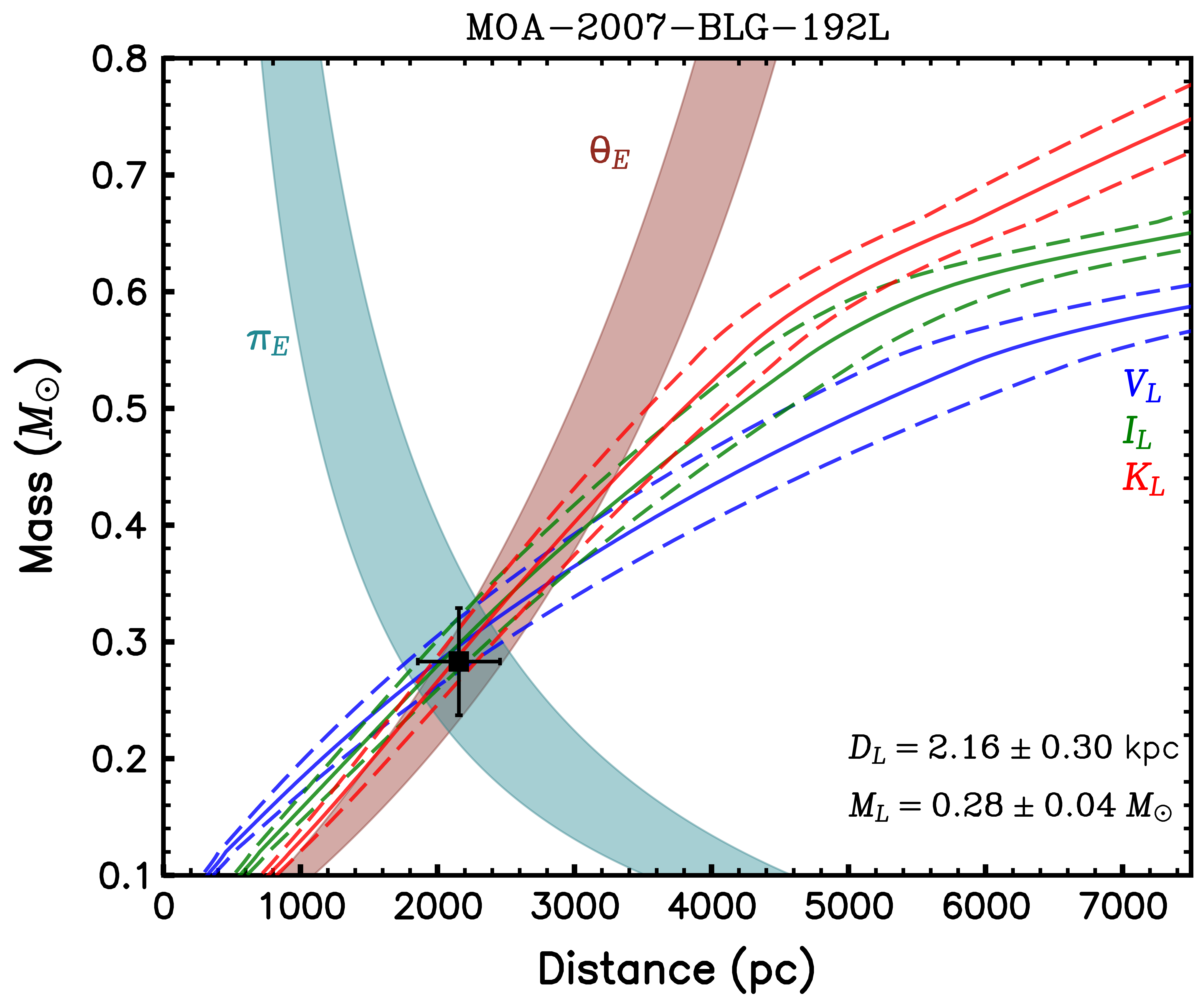}

\centering
\caption{The mass-distance relation for MB07192L with constraints from the lens flux measurement in \textit{HST V} (blue), \textit{HST I} (green), and Keck \textit{K} (red). Dashed lines show the 1$\sigma$ error bars for each passband. The solid teal region shows the mass-distance relation calculated using the microlensing parallax measurement ($\pi_E$), and the solid brown region shows the mass-distance relation calculated using the angular Einstein radius measurement ($\theta_E$).
\label{fig:md_relation}}
\end{figure*}

\indent Additionally, since we have a direct measurement of lens flux in the $V$, $I$, and $K$-bands, we utilize the \cite{delfosse:2000a} empirical mass-luminosity relations in each of these passbands as described by \cite{bennett:2018a}. We consider the foreground extinction in each passband (i.e. Table \ref{table_extinction}) and generate the relations in conjunction with the mass-distance relations given by equations \ref{eq:md_thetaE} and \ref{eq:md_piE}. Figure \ref{fig:md_relation} shows the measured mass and distance of the MB07192 lens. The blue (\textit{HST V}), green (\textit{HST I}), and red (Keck \textit{K}) curves represent the mass-distance relations obtained from the empirical mass-luminosity relations with lens flux measurements given in Table \ref{table:dual-phot}. The dashed lines represent the $1\sigma$ error from the Keck and \textit{HST} measurements. Further, the mass-distance relation obtained from the measurement of $\theta_{E}$ (i.e. Equation \ref{eq:md_thetaE}) is shown as a solid brown region. Considering only these two relations (empirical mass-luminosity and $\theta_E$), there is overlap for a significant amount of mass and distance space. This is sometimes referred to as the ``continuous degeneracy" \citep{gould:2022a}. Fortunately this degeneracy is broken when we include the constraint from the microlensing parallax measurement, $\pi_E$, shown as the solid teal region in Figure \ref{fig:md_relation}.

\indent Table \ref{tab:lightcurve_par} shows the results of the four degenerate light curve models and the Markov chain average for all four models. Although we are able to successfully reduce the number of possible  binary lens solutions presented in K12 by a factor of two, the close/wide and $u_0$ degeneracies still remain. Further, the host star mass is very precisely measured now, however the best-fit mass ratio, $q$, remains largely uncertain because of poor sampling of the light curve. Table \ref{tab:lens-params} gives the derived lens system physical parameters along with their 2$\sigma$ ranges. The large error on the mass ratio results in a large error in the inferred planet mass (see Table \ref{tab:lens-params}).  The lens system properties (host mass, planet mass, etc) shown in Table \ref{tab:lens-params} and Figure \ref{fig:lens_posteriors} are derived from the combined cumulative probability distributions that incorporate the MCMC distributions given by all of the models (e.g. Table \ref{tab:lightcurve_par} columns), weighted by their respective $\chi^2$ fit values.

\indent  Further, the caustic-crossing models are disfavored by a total of $\Delta \chi^2 {\sim} 13$. Since this difference is not particularly large, we include the possible caustic-crossing models in our MCMC sums. However they do not significantly change the overall results, and they have a very low weighting of $e^{\frac{-\chi^2}{2}}=0.0015$. The $\chi^2$ differences are spread across many parameters, some of the largest contributors come from; source magnitudes ($\Delta \chi^2 = 1.19$), source distance ($\Delta \chi^2 = 3.24$), and the photometric light curve fit itself ($\Delta \chi^2 = 7.95$). Lastly, we note that the caustic-crossing models span a relatively small volume in parameter space, which can be clearly seen from Figure 5 of B08. All of these factors contribute to the overall low likelihood for any of the caustic-crossing models in this event. \\
\indent The MB07192 lens system is located at a distance of ${\sim}2.2$ kpc and has a log mass ratio of $log_{10}(q) = -3.87 \pm 0.53$. The host star is directly detected in several high resolution imaging passbands, enabling us to precisely measure its mass to be $M_{\textrm{host}} = 0.28 \pm 0.04M_{\sun}$ with a less-precisely measured mass of the planet to be $m_{\textrm{planet}} = 12.49^{+65.47}_{-8.03}M_{\oplus}$. These masses are consistent with a planet with mass between a super-Earth and sub-Saturn orbiting an M4V dwarf star near the bottom of the main sequence for the redder, foreground disk star population (Figure \ref{fig:CMD}). Figure \ref{fig:lens_posteriors} shows the final posterior probability distributions for the planetary companion mass, host star mass, 2D projected separation, and lens system distance. We note the most likely mass for the planet is in the super-Earth regime (${\sim}3 - 12M_{\oplus}$), as given by the top-left panel in Figure \ref{fig:lens_posteriors}. The best-fit solution gives a 2D projected separation of $a_{\perp} = 2.02 \pm 0.44\,$AU. These physical parameters are calculated from the best-fit solution which takes a combined weighting of several models along with a Galactic model prior based on \cite{koshimoto:2021a}. \\

\begin{deluxetable*}{lccc}[!htp]
\deluxetablecaption{Lens System Properties with Lens Flux Constraints \label{tab:lens-params}}
\tablecolumns{4}
\setlength{\tabcolsep}{14.5pt}
\tablewidth{\columnwidth}
%\tableheight{12pt}
\tablehead{
\colhead{\hspace{-6cm}Parameter} & \colhead{Units} &
\colhead{Values \& RMS} & \colhead{2-$\sigma$ range}
}
\startdata
Angular Einstein Radius ($\theta_E$) & mas & $0.854 \pm 0.043$ & $0.775-0.947$\\
Geocentric lens-source relative proper motion ($\mu_{\textrm{rel,G}}$) & mas/yr & $3.14 \pm 0.15$ & $2.84-3.44$\\
Host Mass ($M_{\rm host}$) & $M_{\Sun}$ & $0.28 \pm 0.04$ & $0.23-0.37$\\
Planet Mass ($m_{\rm p}$) & $M_{\Earth}$ & $12.49^{+65.47}_{-8.03}$ & $2.75-105.06$\\
2D Separation ($a_{\perp}$) & AU & $2.02 \pm 0.44$ & $1.26-2.86$\\
3D Separation ($a_{3\textrm{d}}$) & AU & $2.44^{+1.39}_{-0.68}$ & $1.38-9.65$\\
Lens Distance (D$_{L}$) & kpc & $2.16 \pm 0.30$ & $1.75-2.76$\\
Source Distance (D$_{S}$) & kpc & $7.05 \pm 1.17$ & $4.83-9.38$\\
\enddata
%\tablenotetext{}{\footnotesize{\textbf{Notes.}}}
\end{deluxetable*}

%----------------------------------------------------------Conclusion------------------------------------------------------------------------------------------------------------

\section{Discussion \& Conclusion} \label{sec:conclusion}

\indent Our high resolution follow up observations of the microlensing target MB07192 have allowed us to make a direct measurement of the lens system flux in multiple passbands ($V$, $I$, $K$) as well as a precise determination of the amplitude and direction of the lens-source relative proper motion $\mu_{\textrm{rel}}$. We perform simultaneous multiple-star PSF fitting to obtain best-fit positions and fluxes for both stars across two independent platforms (\textit{HST} and Keck). The lens flux measurements we make enable us to use mass-luminosity relations and new constraints on higher order light curve effects ($\pi_E$, $\theta_E$) to measure a precise mass and distance for the lens system.

\indent Further, we demonstrate the importance of applying constraints from high resolution follow up imaging on the microlensing light curve modeling. Particularly, the microlensing parallax effect, which is present in all microlensing events observed from a heliocentric reference frame, is tightly constrained when the direction of $\mu_{\textrm{rel}}$ can be measured through high resolution imaging. This measurement is critically important for several reasons; poor light curve sampling (i.e. for MB07192) can result in a lack of a microlensing parallax signal from the light curve alone, even for long timescale events. Second, the mass-distance relation that results from a direct measurement of $\pi_E$ (via lens-source separation) allows for the ``continuous degeneracy" to be completely broken.

\begin{figure*}
\centering
\includegraphics[width=0.85\linewidth]{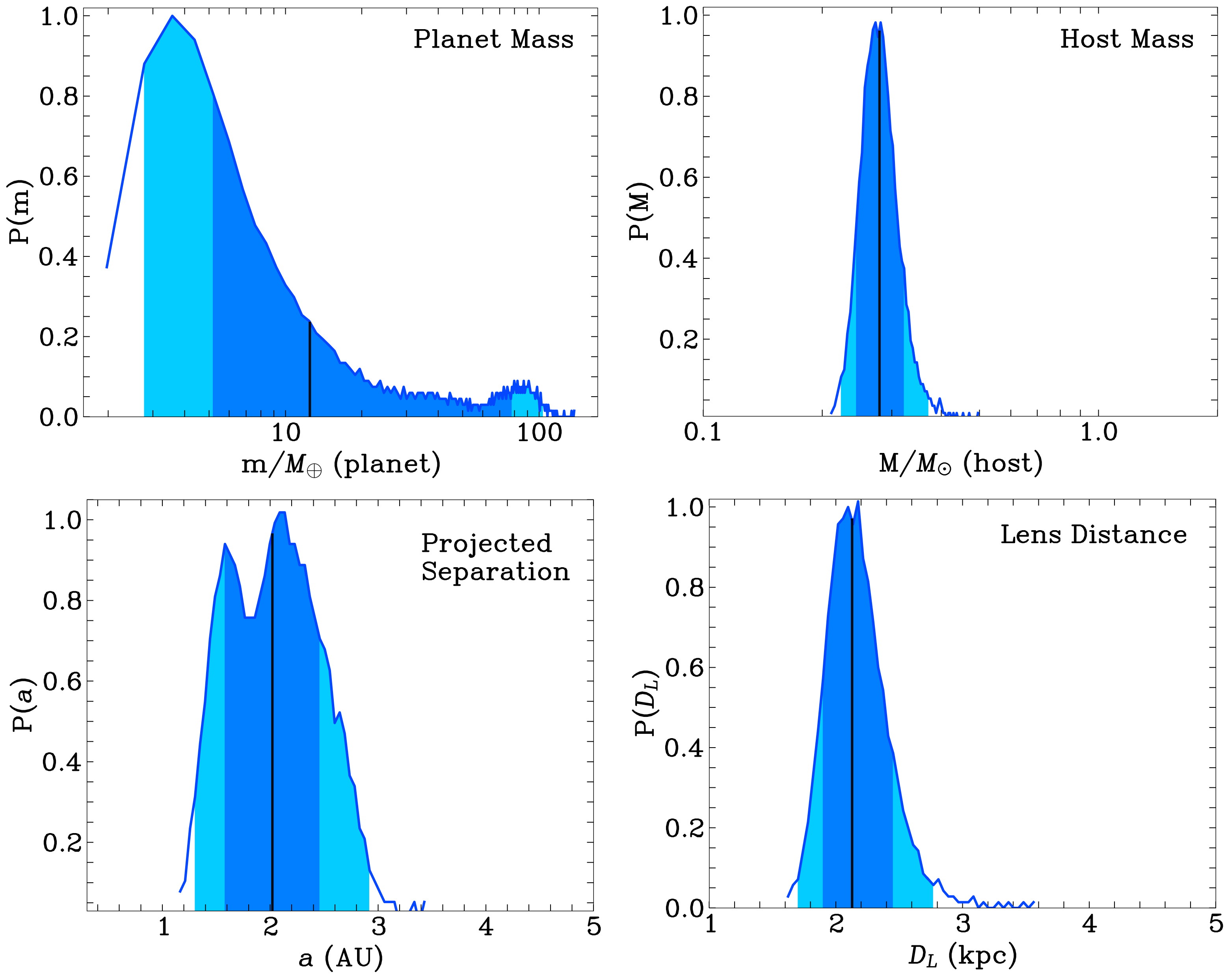}

\centering
\caption{The posterior probability distributions for the lens system physical parameters: planetary companion mass (upper-left), host mass (upper-right), 2D projected separation (lower-left), and lens distance (lower-right). The vertical black line shows the median of the probability distribution for each parameter. The central 68\% of the distributions are shown in dark blue with the remaining central 95\% shown in light blue.
\label{fig:lens_posteriors}}
\end{figure*}

\indent Although the host mass and lens system distance have now been precisely measured as a result of the direct detection in \textit{HST} and Keck, the sampling of the light curve photometry during the microlensing event remains poor. This means that the large uncertainty in the mass ratio parameter, ($q$ in Table \ref{tab:lightcurve_par}), results in a large error in the inferred mass of the planetary companion ($m_{\rm p}$ in Table \ref{tab:lens-params}). As previously mentioned in Section \ref{sec:event}, the large uncertainty in the planetary companion mass comes from a combination of factors; the event is located in a MOA field with a relatively low cadence which leads to poor sampling of the light curve, and the planetary signal was not detected in real time. It was several days after the photometric peak that the anomaly in the light curve was alerted. \\
\indent In conclusion, the distance to the MB07192 lens system is ${\sim}3\times$ larger than previously reported, now at a distance of approximately 2 kpc. Both the mass of the host star and planetary companion are also $2-5\times$ larger than previously reported, which now extends the possible mass range for the planet to a possible sub-Saturn class planet. However, as the top-left panel of Figure \ref{fig:lens_posteriors} shows, the most likely mass for the planetary companion remains in the super-Earth range (${\sim}3 - 12M_{\oplus}$). The previous studies reported a smaller planet mass and also underestimated the error bars on the planet mass for several reasons. All of the B08 models report a too-large microlensing parallax value, which led to a smaller derived planet mass compared to the median value of the planet mass that we report here. Further, the B08 and K12 caustic-crossing models contributed significant weighting to the combined results which gave much smaller error bars on the derived planet mass. Our new results have ruled out the caustic crossing models, which now gives larger error bars on the derived planet mass, particularly the upper 1$\sigma$ error. \\
\indent The results of this work have several implications for the upcoming \textit{RGES}. If \textit{Roman} is expected to employ lens flux measurement methods similar to those described in this work, then a careful selection of secondary observing filters must be made to avoid or minimize instances of the ``continuous degeneracy". For example, the mass-luminosity relation given by a bluer \textit{Roman} passband would have a smaller overlap with the mass-distance relation given by $\theta_E$ compared to other redder \textit{Roman} filters. This effect is more severe for nearby M dwarf lenses (i.e. Figure \ref{fig:md_relation}). Also, for \textit{Roman} detected events with very faint sources or very short Einstein timescales that don't have a measurable microlensing parallax signal, a successful lens-source flux measurement by \textit{Roman} itself will be important for breaking possible degeneracies like those discussed in this work.

%\begin{acknowledgements}
\section*{Acknowledgements}
\indent The authors would like to thank the anonymous referee for helpful comments that improved the structure and led to a stronger manuscript. This paper is based in part on observations made with the NASA/ESA Hubble Space Telescope, which is operated by the Association of Universities for Research in Astronomy, Inc., under NASA contract NAS 5-26555. The Keck Telescope observations and data analysis were supported by a NASA Keck PI Data Award, 80NSSC18K0793, administered by the NASA Exoplanet Science Institute. All of the {\it HST} data used in this paper can be found in MAST: \dataset[10.17909/wbe0-3a21]{http://dx.doi.org/10.17909/wbe0-3a21}. Data presented herein were obtained at the W. M. Keck Observatory from telescope time allocated to the National Aeronautics and Space Administration through the agency's scientific partnership with the California Institute of Technology and the University of California. The Observatory was made possible by the generous financial support of the W. M. Keck Foundation. The authors wish to recognize and acknowledge the very significant cultural role and reverence that the summit of Maunakea has always had within the indigenous Hawaiian community. We are most fortunate to have the opportunity to conduct observations from this mountain. The material presented here is also based upon work supported by NASA under award number 80GSFC21M0002. This work was supported by the University of Tasmania through the UTAS Foundation and the endowed Warren Chair in Astronomy and the ANR COLD-WORLDS (ANR-18-CE31-0002). Part of this work was authored by employees of Caltech/IPAC under Contract No. 80GSFC21R0032 with the National Aeronautics and Space Administration. Lastly, portions of this research were supported by the Australian Government through the Australian Research Council Discovery Program (project number 200101909) grant awarded to A.C. and J.P.B.
%\end{acknowledgements}

\textit{Software}: DAOPHOT-II \citep{stetson:1987a}, daophot$\_$mcmc \citep{terry:2021a}, eesunhong \citep{bennett:1996a}, emcee \citep{foreman:2013a}, genulens \citep{koshimoto:code}, hst1pass \citep{anderson:2022a}, KAI \citep{lu:2022a}, Matplotlib \citep{hunter:2007a}, Numpy \citep{oliphant:2006a}, Spyctres \citep{bachelet:2024a}, SWarp \citep{bertin:2010a}.

\bibliography{main}{}
\bibliographystyle{aasjournal}

\appendix

\section*{2023 \textit{HST} Snapshot Images}
\begin{figure}[!htb]
\centering
\includegraphics[width=0.60\linewidth]{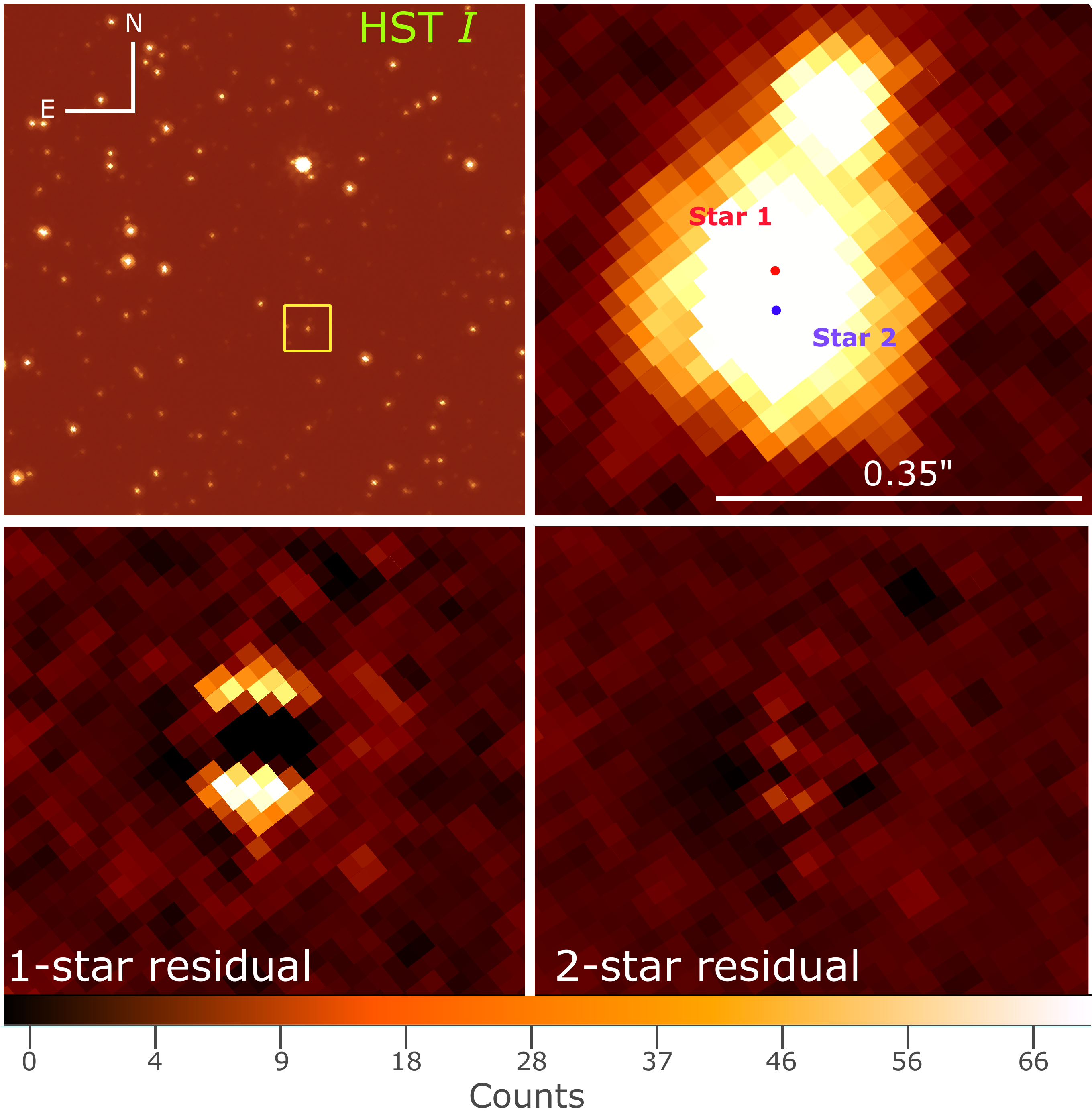}

\centering
\caption{\textit{Top Left:} The 2023 \textit{HST I} band stack image created from two individual exposures from the Snapshot Program. The target is indicated with a yellow box. \textit{Top Right:} Zoomed image of the target, with the two points indicating the best-fit positions for the two stars from the multi-star PSF fitting. We note the unrelated neighbor star has moved closer to the target(s) by ${\sim}1$ pixel between the 2012 and 2023 \textit{HST} data. \textit{Bottom Left:} The residual image from a single-star PSF fit, showing a strong signal of the blended lens(source). \textit{Bottom Right:} Residual image for the simultaneous two-star PSF fit, showing a smoother subtraction with Poisson noise remaining as well as systematics due to the less-characterized PSF model. The color bar represents the pixel intensity (counts) seen in the top-right and bottom-left/right panels. \label{fig:quad-panel_hst_2023}}

\end{figure}

\noindent Figure \ref{fig:quad-panel_hst_2023} shows the four-panel figure created from the two exposures taken during the August 2023 Snapshot Program \citep{sahu:2023prop}. The stacked frame has noticeably larger Poisson noise than the previous \textit{HST} epochs which have 4$\times$ more exposures. The longer time baseline between the peak of the microlensing event and the 2023 \textit{HST} data helps to mitigate the lack of exposures, as the larger separation between source and lens can be clearly detected in this epoch.

\indent This 2023 data, in conjunction with the previous \textit{HST} epochs, largely confirm that the two stars we detect are the true source and lens separating from each other with their expected relative proper motions. This multi-epoch tracking rules out the scenarios in which we are detecting an unrelated blend or a bound stellar companion to either the source or lens.

\section*{Full Light Curve Modeling Comparison}
\begin{figure*}[!htb]
\centering
\includegraphics[width=\linewidth]{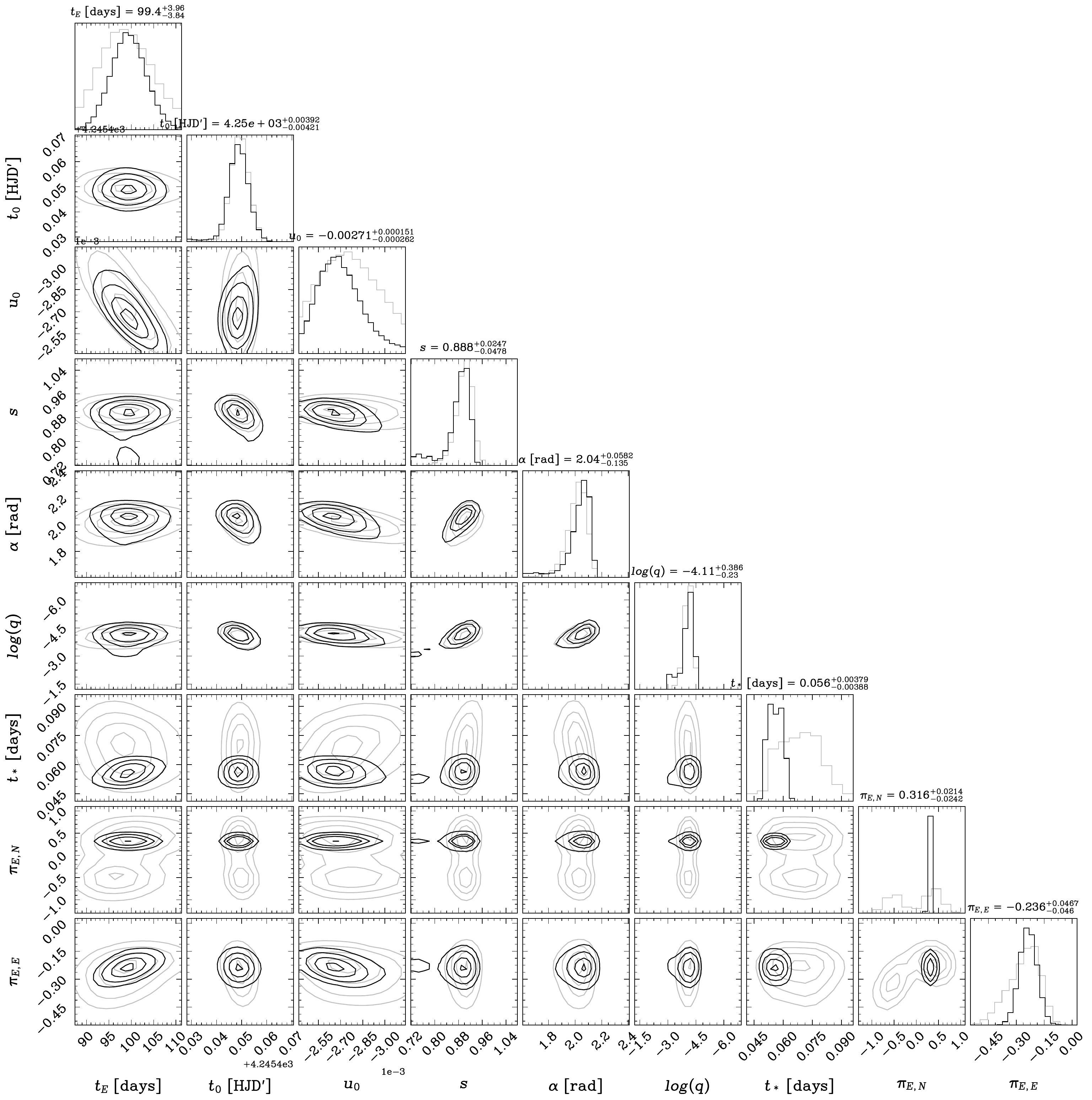}

\centering
\caption{Comparison of model parameter distributions from the light curve photometry only (light grey) and light curve photometry plus \textit{HST}/Keck imaging constraints (black). The two cases shown are for the $s < 1, u_0 < 0$ model. The constraints from the high-resolution imaging are tightened most for the North and East components of the microlensing parallax ($\pi_{E,N}$, $\pi_{E,E}$) as well as the source radius crossing time ($t_*$). The median values given in the title headings (above each histogram) are for the constrained light curve $+$ imaging model. All other model parameters give consistent distributions between the two cases.} \label{fig:model_comparisons}

\end{figure*}

\noindent We show in Figure \ref{fig:model_comparisons} the comparison of the fitting parameters between the light curve modeling from photometry only and from photometry plus \textit{HST}/Keck AO imaging. In Sections \ref{sec:follow-up} and \ref{sec:lens-properties} we explained one of the strongest high-res imaging constraints is that of the microlensing parallax vectors, particularly the North component, $\pi_{\textrm{E,N}}$. Additionally, the tighter constraint on the source radius crossing time, $t_*$, comes primarily from the $\mu_{\textrm{rel,H}}$ measurement derived from the Keck data via the following equation:

\begin{equation} \label{eq:tstar_equation}
    t_* = \frac{\theta_*}{\mu_{\textrm{rel}}}
\end{equation}

\noindent where $\theta_*$ is the angular size of the source star, which we estimate using surface brightness relations from \cite{boyajian:2014a}, considering only stars spanning the range in colors that are relevant for microlensing targets. This yields an angular source size of $\theta_{*} = 0.47 \pm 0.09\ \mu$as for this target. The value of $\mu_{\textrm{rel}}$ in Equation \ref{eq:tstar_equation} comes from the best-fit lens-source separation measurement in Keck.

\indent As described in Section \ref{sec:lens-properties}, the caustic-crossing models are largely ruled out, and the models with a close approach to a caustic cusp do not strongly constrain $t_*$ very well. Ultimately, we can further reduce the total number of possible solutions from K12 (8 solutions) by a factor of 2, which leaves a four-fold degeneracy remaining (i.e. $s \rightarrow 1/s$ and $u_0 < 0$, $u_0 > 0$).

\clearpage

\section*{SED Fitting}
\begin{figure}[!htb]
\centering
\includegraphics[width=0.65\linewidth]{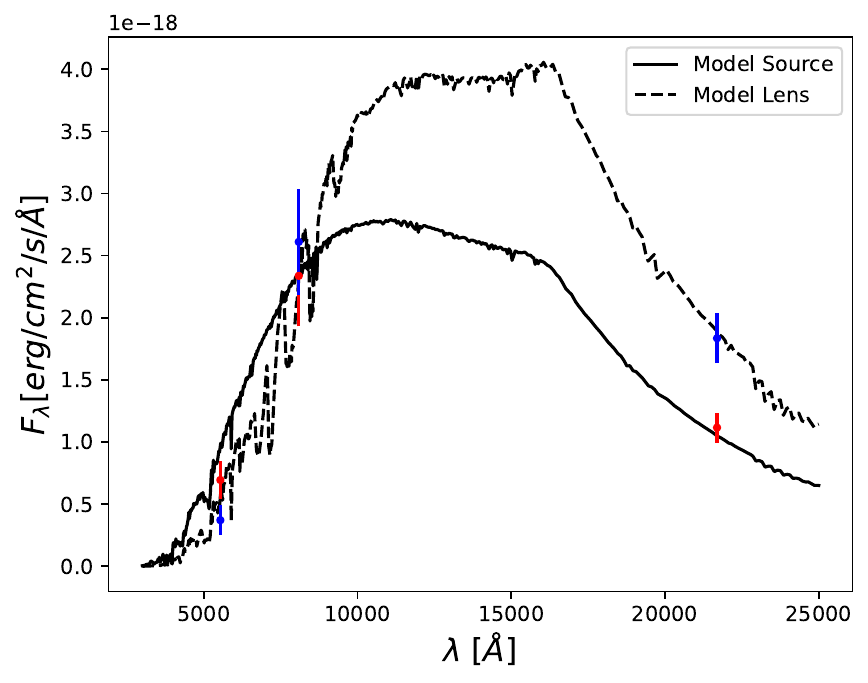}

\centering
\caption{Spectral energy distribution measurements for the source (red) and lens (blue) for MB07192 in the V, I and K bands. The model spectra for the source star (solid line) and lens star (dashed line) are from Kurucz et al. while the absorption law of Wang et al. is adopted.}
\label{fig:SED}
\end{figure}

Using the direct $V$, $I$ and $K-$band magnitude measurements for both the source and the lens stars, it's possible to perform a spectral energy distribution (SED) fitting for these two objects. We used the \texttt{Spyctres} software \citep{bachelet:2024a} to model the stars fluxes, parameterized with $\theta_*$, $T_{\textrm{eff}}$, [Fe/H], and $log(g)$. The spectra template were generated with the \cite{kurucz:1993a} models and the extinction is modeled using the absorption laws from \cite{wang:2019a}, which use only $A_V$ as free parameter. We note that the absorption towards the lens has been parameterized with $\epsilon$ = $A_{V_L}/A_{V_S}$. We use a Gaussian prior on the source extinction $A_V$ from Table 3. The posterior distribution was explored with the Monte-Carlo Markov Chain (MCMC) algorithm implemented in \texttt{emcee} \citep{foreman:2013a}. As can be seen in Figure \ref{fig:SED}, the best models replicate the observations accurately. Lastly, the angular source radius modeled with \texttt{Spyctres} ($\theta_* = 0.49 \pm 0.04\, \mu$as) is in excellent agreement with our earlier estimation based on \cite{boyajian:2014a} ($\theta_* = 0.47 \pm 0.09\, \mu$as).

\clearpage

\section*{Near-Infrared Color-Magnitude Diagram}
We include here the $(J-H, H)$ near-IR CMD for the MB07192 field. Similar to Figure \ref{fig:CMD}, the \textit{HST} CMD of all detected sources from the 2012 epoch is shown in green, with OGLE-III stars cross-identified in the VVV catalog shown in red. Although there is no direct identification of the lens in the \textit{HST} $J$ and $H$-band data, we estimate the lens magnitude via the excess flux (e.g. blend) that is measured on top of the source star in these two passbands. The lens star is estimated to be $(J-H)_L,H_L = [0.98 \pm 0.08, 18.91 \pm 0.15]$, and the source star is $(J-H)_S,H_S = [0.83 \pm 0.05, 19.12 \pm 0.14]$. These estimates are consistent with the source/lens magnitude directly measured in the other passbands (\textit{HST V, I}, Keck $K$), considering the $E(J-K)$ reddening, $A_J$ and $A_H$ extinctions (Table \ref{table_extinction}). \\
\indent Lastly, we show two near-IR isochrones with sub-solar metallicity from the MESA Isochrones \& Stellar Tracks (MIST) database \citep{choi:2016a,paxton:2015a,dotter:2016a}. This includes stars approximately 10 Gyr in age, with metallicity [Fe/H]$=-0.25$ and mass fraction [Z]$=0.01$. The isochrone given by the black curve is well fit to the observed (background) bulge population of stars in the field, and the isochrone given by the grey curve is well fit to the observed (foreground) disk population of stars. As previously mentioned, we deduce the lens is likely an M4 dwarf in the disk at a distance of ${\sim}2$ kpc. The source is likely a type-G main sequence star in the Galactic bulge at a distance of ${\sim}7$ kpc.

\begin{figure}[!htb]
\centering
\includegraphics[width=0.55\linewidth]{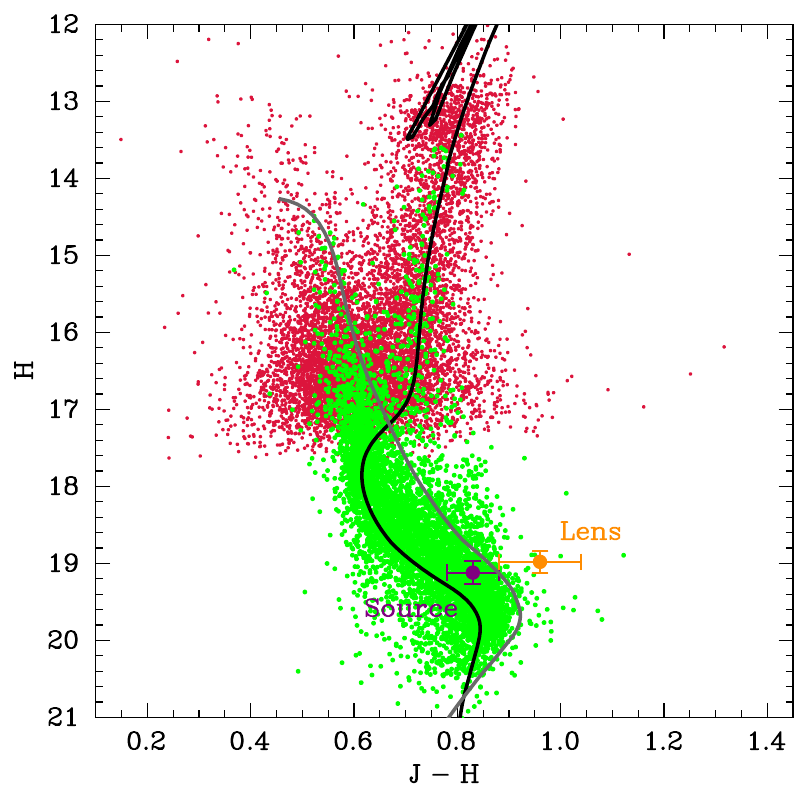}

\centering
\caption{Similar to Figure \ref{fig:CMD}, but for the \textit{HST} near-IR passbands ($F125W-$J band, $F160W-$H-band). The purple and orange points indicate the inferred source and lens colors and magnitudes with associated uncertainties. MIST isochrones for the observed bulge population (black curve) and foreground disk population (grey curve) are shown.}
\label{fig:CMD_IR}
\end{figure}

\end{document}